\documentstyle[epsfig]{mn}

\newif\ifAMStwofonts

\newcommand\lta{\mathrel{\rlap{\lower 3pt\hbox{$\mathchar"218$}}
     \raise 2.0pt\hbox{$\mathchar"13C$}}}
\newcommand\gta{\mathrel{\rlap{\lower 3pt\hbox{$\mathchar"218$}}
     \raise 2.0pt\hbox{$\mathchar"13E$}}}
\newcommand\kms{km~s$^{-1}$}

\newcommand\etal{{et~al.}} 
\newcommand\sigth{\ifmmode \sigma_{\rm th}\else$\sigma_{\rm th}$\fi}
\newcommand\sigv{\ifmmode \sigma_v\else$\sigma_v$\fi}
\newcommand\mM{\ifmmode(m{-}M)\else$(m{-}M)$\fi}
\newcommand\h{\ifmmode H_0\else$H_0$\fi}
\newcommand\msun{\ifmmode{\hbox{M$_\odot$}}\else{M$_\odot$}\fi}

\newcommand\bi{\ifmmode \beta_{I}\else$\beta_I$\fi}
\newcommand\bo{\ifmmode \beta_{O}\else$\beta_O$\fi}

\newcommand\vb{$V$-band}
\newcommand\mgii{Mg$_2$}
\def\ppm{\phantom{$\pm$}}
\def\mi{\ifmmode\overline{m}_I\else$\overline{m}_I$\fi}
\def\mv{\ifmmode\overline{m}_V\else$\overline{m}_V$\fi}
\def\mbar{\ifmmode\overline{m}\else$\overline{m}$\fi}
\def\Mbar{\ifmmode\overline{M}\else$\overline{M}$\fi}
\def\lbar{\ifmmode\overline{L}\else$\overline{L}$\fi}
\def\nbar{\ifmmode\overline{N}\else$\overline{N}$\fi}
\def\Nbar{\ifmmode\overline{N}\else$\overline{N}$\fi}
\def\ibar{\ifmmode\overline{I}\else$\overline{I}$\fi}
\def\vbar{\ifmmode\overline{V}\else$\overline{V}$\fi}
\def\vbib{\ifmmode(\overline{V}{-}\overline{I})\else$(\overline{V}{-}\overline{I})$\fi}
\def\vbkb{\ifmmode(\overline{V}{-}\overline{K})\else$(\overline{V}{-}\overline{K})$\fi}
\def\ibkb{\ifmmode(\overline{I}{-}\overline{K})\else$(\overline{I}{-}\overline{K})$\fi}
\def\ubar{\ifmmode\overline{U}\else$\overline{U}$\fi}
\def\bbar{\ifmmode\overline{B}\else$\overline{B}$\fi}
\def\rbar{\ifmmode\overline{R}\else$\overline{R}$\fi}
\def\jbar{\ifmmode\overline{J}\else$\overline{J}$\fi}
\def\hbar{\ifmmode\overline{H}\else$\overline{H}$\fi}
\def\kbar{\ifmmode\overline{K}\else$\overline{K}$\fi}
\def\Mi{\ifmmode\overline{M}_I\else$\overline{M}_I$\fi}
\def\Miz{\ifmmode\overline{M}_I^0\else$\overline{M}_I^0$\fi}
\def\Mv{\ifmmode\overline{M}_V\else$\overline{M}_V$\fi}
\def\Mvz{\ifmmode\overline{M}_V^0\else$\overline{M}_V^0$\fi}

\def\MbB{\ifmmode\overline{M}_B\else$\overline{M}_B$\fi}
\def\MbV{\ifmmode\overline{M}_V\else$\overline{M}_V$\fi}
\def\MbR{\ifmmode\overline{M}_R\else$\overline{M}_R$\fi}
\def\MbI{\ifmmode\overline{M}_I\else$\overline{M}_I$\fi}
\def\MbJ{\ifmmode\overline{M}_J\else$\overline{M}_J$\fi}
\def\MbH{\ifmmode\overline{M}_H\else$\overline{M}_H$\fi}
\def\MbK{\ifmmode\overline{M}_K\else$\overline{M}_K$\fi}

\def\vi{\ifmmode(V{-}I)\else$(V{-}I)$\fi}
\def\viz{\ifmmode(V{-}I)_0\else$(V{-}I)_0$\fi}
\def\dn{\ifmmode D_n{-}\sigma\else$ D_n{-}\sigma$\fi}
\def\aj{AJ}
\def\apj{ApJ}
\def\apjl{ApJ}
\def\apjs{ApJS}
\def\mnras{MNRAS}

\def\pasp{PASP}
\def\aap{A\&A}

\def\aaps{A\&AS}

\title{Stellar populations and surface brightness fluctuations:
new observations and models}

\author[J. P. Blakeslee et al.]{John P. Blakeslee$^1$, 
Alexandre Vazdekis$^1$, and Edward A. Ajhar$^2$ \\
$^1${Department of Physics, University of Durham, South Road, 
Durham, DH1 3LE, United Kingdom}\\
$^2${Kitt Peak National Observatory, National Optical Astronomy
Observatories, P.\,O. Box 26732, Tucson, AZ 85726, U.S.A.}\\
}

\date{Submitted\,\  10 March 2000.~  Accepted\,\ 8 August 2000.}

\pagerange{\pageref{firstpage}--\pageref{lastpage}}
\pubyear{2000}

\begin{document}

\maketitle
\label{firstpage}

\begin{abstract}
We investigate the use of surface brightness fluctuations (SBF)
measurements in optical and near-IR bandpasses for both stellar
population and distance studies.  New $V$-band SBF data are reported
for five galaxies in the Fornax cluster and combined with
literature data to define a $V$-band SBF distance indicator, calibrated
against Cepheid distances to the Leo group and the Virgo and Fornax
clusters.  The colour dependence of the $V$-band SBF indicator is only
$\sim\,$15\% steeper than that found for the $I$-band, and the mean
`fluctuation colour' of the galaxies is
$\langle\vbar{-}\ibar\rangle\approx2.4$.

We use new stellar population models, based on the latest Padua
isochrones (Girardi \etal\ 2000) transformed empirically to the
observational plane, to predict optical and near-IR SBF magnitudes and
integrated colours for a wide range of population ages and
metallicities.  We examine the sensitivity of the predicted SBF--colour 
relations to changes in the isochrones, stellar transformations, and
initial mass function.
The new models reproduce fairly well the weak dependence of $V$ and $I$ SBF
in globular clusters on metallicity, especially if the more metal-rich
globulars are younger.  Below solar metallicity, the near-IR SBF magnitudes
depend mainly on age while the integrated colours depend mainly on
metallicity.  This could prove a powerful new approach to the 
age-metallicity degeneracy problem; near-IR SBF observations of globular
clusters would be an important test of the models.

The models also help in understanding the \vbib\ and \ibkb\ fluctuation
colours of elliptical galaxies, with much less need for composite stellar
populations than in previous models.  However, in order to
obtain theoretical calibrations of the SBF distance indicators, we combine
the homogeneous population models into composite models and select out
those ones with fluctuation colours consistent with observations.  We are
able to reproduce the observed range of elliptical galaxy \vi\ colours,
the slopes of the $V$ and $I$ SBF distance indicators against \vi\ (fainter
SBF in redder populations), and the flattening of the $I$-band
relation for $\vi\lta1.0$.  The models also match the observed slope of
$I$-band SBF against the Mg$_2$ absorption index and explain the steep
colour dependence found by Ajhar \etal\ (1997) for the {\it HST}/WFPC2
F814W-band SBF measurements.  In contrast to previous models, ours predict
that the near-IR SBF magnitudes will also continue to grow fainter for
redder populations.~~

The theoretical $V$-band SBF zero point predicted by these models agrees
well with the Cepheid-calibrated $V$-band empirical zero~point.
However, the model zero~point is 0.15--0.27\,mag too faint in the
$I$-band and 0.24--0.36\,mag too faint in $K$.  The zero points for the
$I$~band (empirically the best determined) would come into close agreement
if the Cepheid distance scale were revised to agree with the
recent dynamical distance measured to NGC\,4258.
We note that the theoretical SBF calibrations are
sensitive to the uncertain details of stellar evolution 
and conclude that the empirical calibrations remain more secure.  However,
the sensitivity of SBF to these finer details potentially make it a
powerful, relatively unexploited, constraint for stellar evolution and
population synthesis.
\end{abstract}

\section{Introduction}

Attempting to uncover
the mix of stellar ages and metallicities within a composite system such
as an elliptical galaxy can be compared to trying to puzzle out the recipe for
an elaborate gourmet dish on the basis of appearance and aroma alone.
The basic ingredients may be obvious, but without the benefit of
actually biting in, deciding how to prepare and combine them to
reproduce the final product is guesswork.

Historically, most attempts at modeling the stellar populations within
elliptical galaxies have concentrated on reproducing broad-band
photometric colours using various `empirical' or `evolutionary' schemes
(e.g., Spinrad \& Taylor 1971; Faber 1972; Tinsley 1972, 1978; Bruzual 1983)
to combine stars of different masses, metallicities, and/or ages.
More recently, stellar population models have also included predictions for
the absorption line strengths in the integrated spectra 
(e.g., Worthey 1994; Buzzoni 1995; Weiss, Peletier, \& Matteucci 1995;
Borges \etal\ 1995; Vazdekis \etal\ 1996).
The combination of broadband colours and absorption line indices allows
for better constraints on the stellar content of galaxies.  However, the 
spectral indices reveal at least two clear shortcomings of stellar
population synthesis:  model populations of reasonable ages ($\lta 18\,$Gyr)
cannot reproduce the weak Balmer absorption indices observed in the more
metal rich Galactic globular clusters (Cohen, Blakeslee, \& Ryzhov 1998;
Gibson \etal\ 1999; Vazdekis \& Arimoto 1999) and the model metal abundances
do not match the ratios observed in giant ellipticals or metal-rich globulars,
both of which exhibit enhancement of Mg and the other alpha elements
(e.g., Peletier 1989; Worthey, Faber, \& Gonzalez 1992;
Vazdekis \etal\ 1997; Worthey 1998; Cohen \etal\ 1998).  
Moreover, because of the degenerate behaviour of the integrated colours and
indices with respect to age and metallicity variations (see Worthey 1994),
even some of the successes of the models prove ambiguous.
Clearly, new insights beyond those afforded by the usual measures of the
integrated stellar spectrum are needed for unraveling the stellar content
of early-type galaxies.

Observations of surface brightness fluctuations (SBF) provide a
powerful, and hitherto fairly neglected, class of constraints for
stellar population models.  The SBF method is most well known for
studies of the extragalactic distance scale (e.g., Tonry \& Schneider
1988; Tonry, Ajhar, \& Luppino 1990; Tonry \etal\ 1997, hereafter SBF-I;
Lauer \etal\ 1998; Pahre \etal\ 1999) and peculiar velocity field (Tonry
\etal\ 2000a, hereafter SBF-II; Blakeslee \etal\ 1999b).  However,
discussions of optical SBF in the context of stellar populations studies
have been given by Tonry \etal\ (1990), Worthey (1993a,b), Buzzoni
(1993), Ajhar \& Tonry (1994), and Sodemann \& Thomsen (1996).  The
implications of near-infrared SBF observations for distances and stellar
populations have been discussed by Luppino \& Tonry (1993), Pahre \&
Mould (1994), Jensen, Luppino, \& Tonry (1996), and Jensen, Tonry, \& 
Luppino (1998).  Blakeslee, Ajhar, \& Tonry (1999a) give a recent
comprehensive review of the SBF method and its applications.
More recently, Liu, Charlot, \& Graham (2000) have presented 
SBF predictions using a completely updated version of the
Bruzual \& Charlot (1993) stellar population models. 
Their work allows for an up-to-date and independent 
(appearing after this paper was submitted)
comparison to our own.

SBF distances are based on a measurement of the quantity \mbar, the
apparent magnitude of the luminosity-weighted mean luminosity of the
individual stars within a galaxy.  This mean luminosity, called \lbar,
is a well-defined property of any stellar system in any given bandpass.
SBF thus begins to provide some taste of the constituent stars of a
galaxy.  Multiband SBF observations yield distance-independent
`fluctuation colours,' e.g., \hbox{$(\vbar{-}\ibar)$}.  The behaviour
of the absolute SBF magnitudes and fluctuation colours, when plotted for
instance against integrated colours, absorption indices, or each other,
must be reproduced before any population synthesis model can be deemed
fully successful.

In the following section, we present new $V$-band SBF measurements 
for five galaxies in the Fornax cluster.
Most SBF observations to date have been done in the
$I$-band and near-IR bands, as the fluctuations in old stellar populations
are much brighter in these bands and expected to show less scatter.
The only sizable $V$-band SBF data~set so far published has been
by Tonry \etal\ (1990) for galaxies in the Virgo cluster. However,
quantities such as \vbib\ may be useful in stellar population studies,
and because of its compactness, the Fornax cluster is a favourite testbed
for such ideas.
In Sec.~\ref{sec:models}, we describe an updated set of the models of
Vazdekis \etal\ (1996) and present SBF magnitude predictions from these.
We compare these predictions to observations of elliptical galaxies
and globular clusters in Sec.~\ref{sec:disc}.  The comparisons provide
insight into the observations and help in gauging the accuracy of 
theoretical calibrations of the SBF distance method.
In Sec.~\ref{sec:theorydists} we construct composite populations
from our simple population models and use these to derive theoretical
calibrations as a function of \vi.
We also compare the Mg$_2$ and \vi\ indices as SBF calibrators.
Sec.~\ref{sec:nbar} discusses the new Tonry fluctuation number \nbar\
(Tonry \etal\ 2000b, hereafter SBF-IV)
and provides an explanation of its observed properties.
The final section summarizes the major results and conclusions.

\begin{table*}
 \centering
 \begin{minipage}{150mm}
  \caption{Fornax Galaxy Data}
\label{tab:obs}
\newdimen\digitwidth
\setbox0=\hbox{\rm0}
\digitwidth=\wd0
\catcode`?=\active
\def?{\kern\digitwidth}
  \begin{tabular}{cccccccccccc}
\hline
Galaxy & $v_h$ & $A_V$ & sec$\,z$ & Exp. & PSF & $m_1^*$ & $V{-}I$ & \vbar & $\pm$ & $\vbar{-}\ibar$ & $\pm$ \\
   & (km/s) & (mag) &   & (sec) & (\arcsec) &(mag) &(mag) & (mag) & (mag) & (mag) & (mag)\\
\hline                                                
N1316 &   1760 & 0.07 & 1.01 & 3000 & 1.04 & 35.22 & 1.13 & 32.25 & 0.16 & 2.42 & 0.22 \\
N1344 &   1169 & 0.06 & 1.02 & 2400 & 1.04 & 34.98 & 1.14 & 31.95 & 0.11 & 2.28 & 0.31 \\
N1380 &   1877 & 0.06 & 1.00 & 2400 & 1.03 & 34.99 & 1.20 & 32.33 & 0.12 & 2.54 & 0.19 \\
N1399 &   1425 & 0.04 & 1.10 & 2400 & 0.94 & 34.98 & 1.23 & 32.47 & 0.12 & 2.36 & 0.18 \\
N1404 &   1947 & 0.04 & 1.02 & 2400 & 1.04 & 35.00 & 1.22 & 32.48 & 0.12 & 2.28 & 0.20 \\
\hline
\end{tabular}
\end{minipage}
\end{table*}

\section{New $V$-band Fornax Data}
\label{sec:obs}

\subsection{Observations and Reductions}

We observed five Fornax cluster galaxies in 1995~August with
the Tek~4 CCD camera at the Cassegrain focus on the 4$\,$m telescope
at Cerro Tololo Inter-American Observatory. 
These data were previously reported on by Blakeslee \& Tonry (1996),
who studied the globular cluster populations of these galaxies.
Four or five 600\,s $V$-band exposures were taken of each galaxy.
The image scale was 0\farcs158~pix$^{-1}$, and the usable area was
about 5\farcm1 square.  
The sky was photometric, and the seeing was $\sim\,$1\arcsec.
The photometry was calibrated using Landolt (1992) standard stars.  

The images were processed and reduced after the standard manner of the
SBF Survey, as described by Tonry \etal\ (1990; SBF-I) and Blakeslee
\etal\ (1999a).  We summarize it here.  Individual data frames are
bias-subtracted, flattened, shifted into registration, and combined,
rejecting pixels affected by cosmic ray hits.  We then fit and remove a
smooth galaxy model and run {\sc DoPhot} (Schechter, Mateo, \& Saha
1993) on the resulting image to detect the (unresolved) globular
clusters, faint background galaxies and Galactic stars.  The detected
objects (ranging from 1650 to 3200 for these images) are then used to
construct a composite luminosity function that gives an estimate for the
variance in surface brightness from faint, undetected sources.  The
detected sources brighter than a cutoff signal-to-noise threshold of 4.5
are masked out of the image, and the amplitude of the surface brightness
fluctuations is derived from the power spectrum in different regions
within the image.

The variance from globular clusters and background galaxies becomes
more important relative to the stellar fluctuations (dominated by
red giant stars) in bluer bands.  However, these images are deep
enough to allow for good characterization of the globular
cluster luminosity functions (GCLFs) of these galaxies, with
the detection completeness going more than a magnitude beyond
the peak in the GCLF (Blakeslee \& Tonry 1996).  
Our GCLF analysis is consistent with the {\it HST} results obtained
in the $B$ and $I$ bands for three galaxies in common (Grillmair
\etal\ 1999; Ferrarese \etal\ 2000 give a complete tabulation).  
Moreover, changing our signal-to-noise cutoff from 4.5 to 5.0 
only changes the final \vb\ SBF magnitudes in random fashion
and at the 0.02\,mag level.

Because of the importance of the
template star used in modeling the power spectrum of the 
point spread function (PSF), we
carried out the SBF reductions with 2 or 3 different stars
in each image except that of NGC\,1380, for which there is only
a single good PSF star.  The differences in \vbar\ calculated
from the different templates were typically $\lta\,0.05$\,mag,
the standard allowance made in the SBF Survey reductions for
PSF mismatch, which we also include in the uncertainty.

Table~\ref{tab:obs} lists for the present data sample:
the galaxy name; heliocentric velocity from 
NED\footnote{The NASA/IPAC Extragalactic Database (NED) is operated
by the Jet Propulsion Laboratory, Caltech, under
contract with the National Aeronautics and Space Administration};
$V$-band Galactic extinction from 
Schlegel, Finkbeiner, \& Davis (1998, hereafter SFD);
mean sec$\,z$ airmass of the observation; exposure time;
full-width at half maximum of the PSF;
magnitude $m_1^*$ for a source yielding 1 photoelectron per
total integration time; mean \vi\ colour of the galaxy from
the SBF Survey catalogue (SBF-IV); $V$-band SBF magnitude \vbar\ 
measured in the current data set; and the \vbib\ fluctuation colour
using the \ibar\ from the SBF Survey catalogue.
The magnitudes and colours are corrected for Galactic extinction.
Taking the difference $\vbar{-}m_1^*$ shows that
at least 10 electrons are collected for each giant star of 
magnitude \vbar, which is above
the average for the $I$-band SBF survey (Blakeslee \etal\ 1999a; SBF-IV),
despite the fluctuations being significantly brighter in $I$.
The darkness of the sky in $V$ adds further
to the advantage of these data.

We find a mean SBF colour 
$\langle \vbar{-}\ibar\rangle_{\rm For} = 2.39\pm0.09$
with an rms dispersion of 0.11\,mag for these Fornax
galaxies.  The dispersion is
smaller than the quoted uncertainties in
Table~\ref{tab:obs}, which yield a reduced $\chi^2$ of
only 0.28.  However, with only 4 degrees of
freedom, the probability of $\chi^2_\nu$ this small is 11\%,
not unreasonable.  
For the twelve Virgo and Leo galaxies with \vbar\ measured by 
Tonry \etal\ (1990), 
updated to use SFD extinctions and to include the same 0.05\,mag
allotment for PSF mismatch, we derive
$\langle \vbar{-}\ibar\rangle_{\rm Vir} = 2.42\pm0.05$,
with a dispersion of 0.15\,mag and  $\chi^2_\nu = 1.09$.
We conclude that over the limited range of stellar populations
found among the early-type galaxies in Virgo and Fornax,
\vbar\ apparently follows the same behaviour as \ibar,
so that \vbib\ is observed to be approximately constant.
Given this, \vbar\ should also work as a distance indicator
for early-type galaxies.

\subsection{The Empirical $V$-band SBF Distance Indicator}
\label{ssec:vemp} 

The small scatter in \vbib\ suggests that the $V$-band absolute
SBF magnitude \Mv\ can be written as
a linear function of \vi, as is the case for \Mi\ (SBF-I).
We choose the same fiducial colour as for $I$ and write~ 
\begin{equation}
\Mv \,=\, \Mvz + \beta\,[\vi - 1.15] \,,
\label{eq:gencal}
\end{equation}
where $\beta$ and \Mvz\ are the slope and zero point of this proposed
universal relation.  Doing a bivariate fit to the Fornax \vbar--\vi\
relation (including an allowance for group depth, as in SBF-I) gives a
slope $\beta_{\rm For} = 4.8\pm1.7$, and the Virgo data yield $5.29\pm0.91$.
Tonry \etal\ (1990) also measured \vbar\ for two Leo ellipticals
(finding $\vbar = 30.61$ for NGC\,3377 although it did not appear
in their tables), and these give a slope of $7.1\pm2.4$.
The weighted average for these three groups is $\beta = 5.37\pm0.76$.
Shifting the three groups together according to their average
Cepheid distances (Ferrarese \etal\ 2000) and doing a single
bivariate fit to the \vbar-\vi\ relation gives a consistent 
slope of $\beta = 5.24\pm0.74$.
If we instead shift the three groups together according to their
$I$-band SBF relative distances (SBF-II)
which are based on many more galaxies, we find $\beta = 5.28\pm0.74$.

The $I$-band SBF measurements should give a more accurate set of
relative distances to the early-type galaxies in these groups,
since the late-type host galaxies of Cepheids are less strongly
clustered (see Ferrarese \etal\ 2000; SBF-II).  However, the
difference is unimportant here.
We adopt $\beta = 5.3\pm0.8$ for the slope of the \vbar-\vi\ relation.
Mindful of possible systematic effects, we have chosen to be conservative
with the errorbar.  For instance, substructure in the Virgo cluster
could make the allowance for an rms group depth a poor approximation.  
In addition, the giant galaxy NGC\,1316 in Fornax is much
more dusty and disturbed than any of the others and may be
a poor choice as a calibrator, and with Leo having
only two galaxies in the sample, there really 
is not much leverage at $\vi \lta 1.13$. 

%
%
\begin{figure}
\epsfig{figure=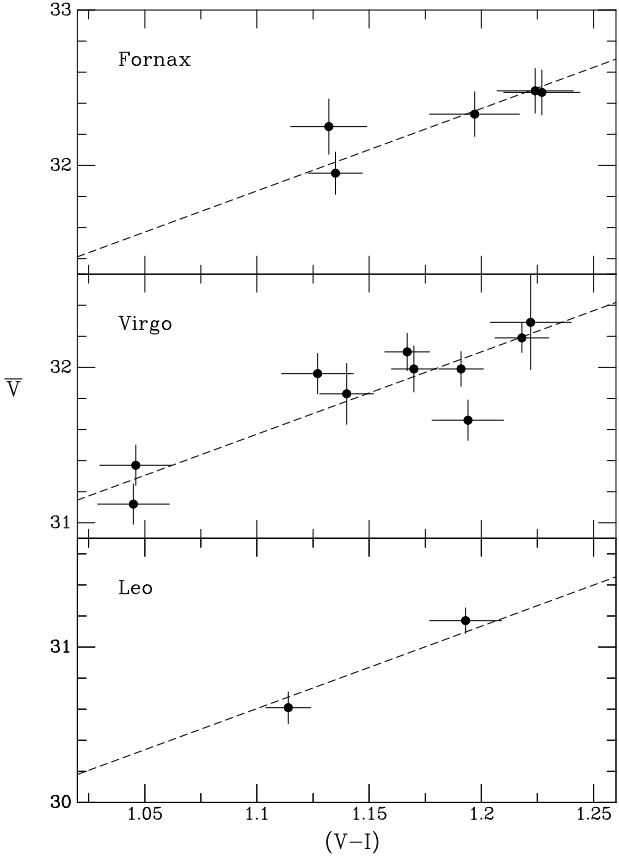, height=0.64\textwidth, width=0.46\textwidth}
\medskip
\caption{The $V$-band SBF vs.\ integrated $(V{-}I)$ colour (both corrected
for Galactic extinction) relation for galaxies in 
the Fornax and Virgo clusters and the Leo group.  The Fornax
data are from the present study and the other data are from
Tonry \etal\ (1990).   The \vbar\ errors have been increased in
quadrature by 0.02, 0.08, and 0.05\,mag for Leo, Virgo, and Fornax,
respectively, to reflect the expected rms group depths.
The integrated colours are from the $I$-band SBF Survey catalogue and
are usually the average of 2 or more photometric observations.  The
dashed lines show fits with the adopted slope of 5.3 (see text).
}\label{fig:vbcal}
\end{figure}

Figure~\ref{fig:vbcal} shows observed \vbar\ plotted against \vi\ for
galaxies in Leo, Virgo, and Fornax.  The data are all from the present
study and Tonry \etal\ (1990).  Fitting with a fixed slope of 5.3 gives
zero points at $\vi=1.15$ of $\vbar^0 = 30.87\pm0.08$, $31.83\pm0.05$,
and $32.10\pm07$ mag for Leo, Virgo, and Fornax, respectively.  Assuming
a universal \vbar-\vi\ relation therefore indicates that Virgo and Fornax
are $0.96\pm0.10$ mag and $1.23\pm0.11$ mag more distant than Leo, respectively.
The Cepheid measurements give distance moduli for Virgo and Fornax with
respect to Leo of $0.95\pm0.08$ and $1.34\pm0.08$ (Ferrarese \etal\ 2000,
excluding the more distant NGC\,1425 from Fornax), and the corresponding
relative moduli from $I$-band SBF are $0.93\pm0.05$ and $1.30\pm0.05$
(SBF-II).  Thus, these three measures (Cepheids, $I$~SBF, and $V$~SBF) of
the relative distances are all consistent.

We conclude that the data support a universal relation as in
Eq.~(\ref{eq:gencal}), although they are not of sufficient quantity for
a very rigorous test.  An absolute calibration for $V$-band SBF can be
obtained from the $I$-band distances calibrated against the Cepheid
distances to groups.
%
%
Calibrated this way, the $I$-band zero point is
$\Miz = -1.62$ mag, which places Leo, Virgo, and Fornax
at $\mM=30.10$, 31.03, and 31.40\,mag.
The resulting $V$ zero point is
$\Mvz = 0.77\pm0.05\pm0.08$ mag, where the first errorbar is the internal
uncertainty and the second errorbar is the estimated systematic
uncertainty in tying SBF groups to the Cepheid distance scale (SBF-II).
(The uncertainty in the Cepheid scale itself is about 0.17\,mag 
according to Mould \etal\ 2000.)  
The fully empirical calibration for $V$-band SBF is then
\begin{equation}
\Mv \,=\, (0.77\pm0.12) \,+\, (5.3\pm0.8)\,[\vi - 1.15] \,.
\label{eq:vbempcal}
\end{equation}
Simply adopting the mean Cepheid
distances measured for spirals in these three groups gives a
virtually identical zero point of $\Mvz = 0.76\pm0.06$ mag.
However, using the $I$-band survey distances calibrated from
our attempts to measure SBF in the bulges of Cepheid-bearing
spirals gives a zero point 0.12\,mag brighter.
We have used the group-calibrated distances because of the close
agreement with the Cepheids distances and because of the large
systematic and random
uncertainties inherent in SBF distances to spirals,
but we have adopted the 0.12 mag offset as 
an estimate of the uncertainty in Eq.~(\ref{eq:vbempcal}).
SBF-II discusses the difficulties in the zero-point
calibration in greater detail.

\section{Stellar Population Models}
\label{sec:models}

\subsection{Description of the Models}

For this investigation we use
the evolutionary stellar population synthesis 
models of Vazdekis \etal\ (1996), as updated by 
Vazdekis (2000). These models calculate colours, line-strengths, 
and mass-luminosity ratios for intermediate-age and old
single-metallicity stellar populations (SSPs). High resolution 
spectral energy distributions for these models 
in selected regions of the optical spectrum
are presented by Vazdekis (1999). The reader is referred to these papers 
for a complete description of the models, but it is important to 
summarize the main ingredients here.

The Vazdekis \etal\ (1996) code made use of the homogeneous set of
Bertelli \etal\ (1994) (i.e., `old Padua') isochrones, extended to
masses M$<$0.6~\msun\ using the stellar tracks of Pols \etal\ (1995),
while here we switch to the Girardi \etal\ (2000) (i.e., `new
Padua') isochrones. The two sets of isochrones cover a wide range of
ages and metallicities and include the latest stages of the stellar
evolution through the thermally pulsing asymptotic giant branch (AGB).
The low-mass cutoff of the
new Padua isochrones has been extended down to 0.15~\msun, and the
largest metallicity computed is Z=0.03 (instead of Z=0.05 as in the old set).
Other differences with respect to the old set of isochrones include an
improved equation of state, updated low-temperature opacities, a
different prescription for the helium fraction $Y$, and a milder amount
of convective overshoot (the extent to which the mean free path of the
convective elements takes them beyond the point at which their
acceleration vanishes).  The AGB treatment has also been revised.
Girardi \etal\ supply more details and references, but as the accumulated
differences in the isochrones produce nonnegligible changes in our own
model results below, we feel it is important to provide some explanation.

The convective mixing length
and the amount of overshoot at the boundaries of the core and envelope
convective zones were tuned so that the 4.6\,Gyr solar metallicity,
solar mass track would match the observed solar radius and luminosity as
well as the helioseismological determination of the convective envelope
depth.  Overshooting in effect increases the size of the convective
zones, which increases the luminosity. In the core it also helps
replenish the hydrogen supply, thus increasing the main sequence 
lifetime (e.g., Bertelli, Bressan, \& Chiosi 1985).  The mixing length
parameter optimized for the solar model was then adopted for all the
stellar models, and the parameters governing convective overshoot were
made appropriate functions of the mass.

The helium fraction was also tuned to the solar model. The new relation
corresponds to $Y\approx 0.23 + 2.25\,Z$, whereas Bertelli \etal\ used a
relation with a slope of 2.5, which they quoted as a lower limit. 
The isochrones used by Worthey (1994) employed a steeper slope of 2.7,
but the solar metallicity definition was 10\% lower.  
We emphasize that this slope is not well known, and there may well
be other factors governing the helium content 
(e.g., Alonso \etal\ 1997).
The helium fraction is important because it
controls the molecular weight, and thus the mass-luminosity relation and
lifetimes of metal-rich stars.  For a fixed mass, an increase of 0.05 in
$Y$ at solar metallicity decreases the lifetime by 40\% (Charlot,
Worthey, \& Bressan 1996), which significantly changes the temperature
difference between the main sequence turnoff and the giant branch.

We caution that since the stellar physics has been specifically tuned to
match the observed properties of the sun, the accuracy depends on our
detailed understanding of the solar interior, as well as the extent to
which the sun is a typical star.  Of course, while the solar neutrino
problem remains unsolved, we must continue to expect imperfect agreement
between theory and observa\-tions.  Moreover, in comparison to solar
ratios, elliptical galaxies (our main concern here) have enhanced
alpha-element abundances with respect to iron-peak abundances.  Alpha
enhancement makes the isochrones cooler, by hundreds of~K, from the main
sequence through the \hbox{RGB} (e.g., Salaris, Chieffi, \& Straniero 1993;
Worthey 1998), although this may simply mimic a global metallicity change.
Another effect, elemental diffusion, makes the turnoff temperature cooler
without much change to the RGB (Straniero, Chieffi, \& Limongi 1997).
However, neither of these effects is understood in detail, and the Padua
isochrones make no allowance for them.

Girardi \etal\ calculate the stellar tracks only to the beginning 
of the thermally pulsing AGB.  For the isochrones, they then employ
a `synthetic' prescription to complete the evolution through this
phase to the point of complete envelope ejection.  
The prescription follows Vassiliadis \& Wood (1993), 
Groenewegen \& de Jong (1993), and Girardi \& Bertelli (1998).
The Padua group is currently further revising their AGB treatment 
to use more detailed models.

The Vazdekis (2000) code transforms the theoretical parameters
(luminosity and temperature) of the isochrones to the observational
plane (e.g., fluxes, colours) following empirical relations inferred
from extensive observational photometric and spectral stellar libraries
rather than using model stellar atmospheres.  The model uses the
metallicity-dependent empirical relations of Alonso \etal\ (1996) for
dwarfs and the ones of Alonso \etal\ (1999) for giants of all
metallicities (each sample is composed of $\sim$500 stars).  A
semi-empirical approach is performed for giants with temperatures cooler
than $\sim$3500~K and dwarfs cooler than $\sim$4000~K on the basis of
the empirical colour-T$_{\rm eff}$ relations of Lejeune \etal\ (1997)
and Lejeune \etal\ (1998), respectively, for solar metallicity; these
are applied to other metallicities according to the
stellar model atmospheres of Hauschildt \etal\ (1999).  This treatment
differentiates this model from most other evolutionary codes (e.g.,
Worthey 1994; Tantalo \etal\ 1996; Kodama \& Arimoto 1997; Kurth \etal\
1999) which are primarily based on theoretical stellar atmospheres, from
which the colours are calculated.  
The bolometric correction we adopt for the sun is
$\hbox{BC}_\odot = -0.12$ with a bolometric magnitude of 4.70.

Absorption line strengths on the Lick/IDS system are also
calculated in our models using the empirical fitting functions of
Worthey \etal\ (1994) and Worthey \& Ottaviani (1997).  These line
indices, as well as the ones of Rose (1994), can also be measured on the
Vazdekis (1999) model spectral energy distributions.

The code assumes two shapes for the initial mass function (IMF):
unimodal, having a power law form with the 
logarithmic slope $\mu$ as a free parameter 
(where $\mu=1.3$ corresponds to a Salpeter [1955] IMF),
and bimodal, which is similar to the previous one but with a transition
to a shallower slope starting at 0.6$\,\msun$, becoming flat for
masses $< 0.4$~\msun (see Vazdekis \etal\ 1996).
The bimodal form can be made to approximate a Miller-Scalo (1979) 
type IMF.

In order to gauge the uncertainty in the SBF predictions presented below,
we have also computed an additional set of models on the 
basis of the isochrones used by Vazdekis \etal\ (1996) (old Padua 
extended to lower masses by the
Pols \etal\ tracks), although applying the new transformations 
to the observational plane. 


\begin{table*}
\centering \begin{minipage}{150mm}
\caption{SBF Predictions from New Models for Salpeter $\mu{=}1.3$ IMF}\label{tab:sbf_un1.3}
\newdimen\digitwidth
\setbox0=\hbox{\rm0}
\digitwidth=\wd0
\catcode`?=\active
\def?{\kern\digitwidth}
\begin{tabular}{crccccrrrrrrrr}
\hline
[Fe/H] & Gyr & $B{-}V$ & $V{-}I$ & $J{-}K$ & Mg$_2$ & $\overline M_U$ & $\overline M_B$ & $\overline M_V$ & 
$\overline M_R$ & $\overline M_I$ & $\overline M_J$ & $\overline M_H$ & $\overline M_K$\\
\hline
$-$1.7 &  4.0 &  0.58 &  0.82 &  0.62 &  
0.046\rlap{\footnote[1]{These numbers are uncertain because they require
extrapolation of the Worthey \etal\ (1994) fitting functions.}} &  
	0.86 &$-$0.26 &$-$2.05 &$-$3.03 &$-$3.88 &$-$5.01 &$-$5.95 &$-$6.16 \\
$-$1.7 &  5.0 &  0.59 &  0.83 &  0.61 &  0.048\rlap{$^a$} &  0.95 &$-$0.12 &$-$1.86 &$-$2.81 &$-$3.64 &$-$4.76 &$-$5.69 &$-$5.90 \\
$-$1.7 &  6.3 &  0.61 &  0.85 &  0.61 &  0.050\rlap{$^a$} &  1.03 &$-$0.01 &$-$1.68 &$-$2.60 &$-$3.41 &$-$4.51 &$-$5.43 &$-$5.63 \\
$-$1.7 &  7.9 &  0.63 &  0.85 &  0.60 &  0.053\rlap{$^a$} &  1.09 &  0.10 &$-$1.50 &$-$2.39 &$-$3.17 &$-$4.24 &$-$5.13 &$-$5.33 \\
$-$1.7 & 10.0 &  0.64 &  0.86 &  0.59 &  0.054 &  1.17 &  0.24 &$-$1.28 &$-$2.13 &$-$2.88 &$-$3.92 &$-$4.78 &$-$4.97 \\
$-$1.7 & 11.2 &  0.65 &  0.86 &  0.59 &  0.056 &  1.19 &  0.29 &$-$1.19 &$-$2.02 &$-$2.76 &$-$3.77 &$-$4.62 &$-$4.81 \\
$-$1.7 & 12.6 &  0.65 &  0.86 &  0.58 &  0.058 &  1.22 &  0.36 &$-$1.09 &$-$1.90 &$-$2.62 &$-$3.62 &$-$4.45 &$-$4.63 \\
$-$1.7 & 14.1 &  0.65 &  0.86 &  0.58 &  0.055 &  1.25 &  0.43 &$-$0.98 &$-$1.77 &$-$2.47 &$-$3.45 &$-$4.27 &$-$4.44 \\
$-$1.7 & 15.8 &  0.64 &  0.85 &  0.58 &  0.058 &  1.27 &  0.50 &$-$0.86 &$-$1.64 &$-$2.32 &$-$3.28 &$-$4.07 &$-$4.24 \\
$-$1.7 & 17.8 &  0.63 &  0.85 &  0.58 &  0.060 &  1.27 &  0.55 &$-$0.78 &$-$1.53 &$-$2.20 &$-$3.13 &$-$3.90 &$-$4.06 \\
$-$1.3 &  4.0 &  0.62 &  0.88 &  0.68 &  0.074\rlap{$^a$} &  1.50 &  0.32 &$-$1.53 &$-$2.62 &$-$3.60 &$-$4.95 &$-$6.02 &$-$6.27 \\
$-$1.3 &  5.0 &  0.65 &  0.89 &  0.68 &  0.077\rlap{$^a$} &  1.59 &  0.42 &$-$1.39 &$-$2.44 &$-$3.41 &$-$4.76 &$-$5.82 &$-$6.07 \\
$-$1.3 &  6.3 &  0.67 &  0.90 &  0.67 &  0.079\rlap{$^a$} &  1.65 &  0.50 &$-$1.26 &$-$2.29 &$-$3.22 &$-$4.53 &$-$5.57 &$-$5.82 \\
$-$1.3 &  7.9 &  0.68 &  0.90 &  0.65 &  0.082\rlap{$^a$} &  1.74 &  0.61 &$-$1.09 &$-$2.08 &$-$2.97 &$-$4.23 &$-$5.25 &$-$5.49 \\
$-$1.3 & 10.0 &  0.70 &  0.91 &  0.65 &  0.083 &  1.81 &  0.71 &$-$0.93 &$-$1.88 &$-$2.73 &$-$3.94 &$-$4.92 &$-$5.15 \\
$-$1.3 & 11.2 &  0.71 &  0.91 &  0.64 &  0.082 &  1.82 &  0.75 &$-$0.87 &$-$1.80 &$-$2.62 &$-$3.79 &$-$4.76 &$-$4.98 \\
$-$1.3 & 12.6 &  0.71 &  0.92 &  0.64 &  0.086 &  1.84 &  0.79 &$-$0.81 &$-$1.71 &$-$2.52 &$-$3.65 &$-$4.60 &$-$4.81 \\
$-$1.3 & 14.1 &  0.72 &  0.92 &  0.63 &  0.088 &  1.84 &  0.82 &$-$0.75 &$-$1.64 &$-$2.43 &$-$3.54 &$-$4.46 &$-$4.67 \\
$-$1.3 & 15.8 &  0.71 &  0.92 &  0.63 &  0.093 &  1.91 &  0.92 &$-$0.61 &$-$1.48 &$-$2.24 &$-$3.31 &$-$4.21 &$-$4.40 \\
$-$1.3 & 17.8 &  0.70 &  0.91 &  0.63 &  0.094 &  1.83 &  0.96 &$-$0.53 &$-$1.38 &$-$2.13 &$-$3.18 &$-$4.06 &$-$4.24 \\
$-$0.7 &  4.0 &  0.73 &  0.99 &  0.77 &  0.124\rlap{$^a$} &  2.59 &  1.39 &$-$0.36 &$-$1.47 &$-$2.83 &$-$4.75 &$-$5.95 &$-$6.17 \\
$-$0.7 &  5.0 &  0.75 &  1.00 &  0.76 &  0.127\rlap{$^a$} &  2.72 &  1.49 &$-$0.26 &$-$1.35 &$-$2.68 &$-$4.57 &$-$5.76 &$-$5.98 \\
$-$0.7 &  6.3 &  0.77 &  1.01 &  0.76 &  0.133 &  2.85 &  1.59 &$-$0.15 &$-$1.23 &$-$2.53 &$-$4.38 &$-$5.57 &$-$5.79 \\
$-$0.7 &  7.9 &  0.79 &  1.03 &  0.76 &  0.144 &  2.95 &  1.68 &$-$0.05 &$-$1.13 &$-$2.40 &$-$4.21 &$-$5.39 &$-$5.61 \\
$-$0.7 & 10.0 &  0.82 &  1.04 &  0.76 &  0.155 &  3.00 &  1.74 &  0.03 &$-$1.04 &$-$2.28 &$-$4.02 &$-$5.19 &$-$5.42 \\
$-$0.7 & 11.2 &  0.83 &  1.05 &  0.76 &  0.161 &  3.00 &  1.76 &  0.06 &$-$0.99 &$-$2.22 &$-$3.91 &$-$5.07 &$-$5.31 \\
$-$0.7 & 12.6 &  0.84 &  1.06 &  0.76 &  0.167 &  2.99 &  1.78 &  0.09 &$-$0.95 &$-$2.15 &$-$3.78 &$-$4.93 &$-$5.18 \\
$-$0.7 & 14.1 &  0.85 &  1.07 &  0.76 &  0.172 &  2.96 &  1.80 &  0.13 &$-$0.91 &$-$2.09 &$-$3.66 &$-$4.80 &$-$5.06 \\
$-$0.7 & 15.8 &  0.86 &  1.07 &  0.76 &  0.177 &  2.91 &  1.83 &  0.18 &$-$0.85 &$-$2.00 &$-$3.50 &$-$4.64 &$-$4.90 \\
$-$0.7 & 17.8 &  0.86 &  1.07 &  0.75 &  0.183 &  2.80 &  1.83 &  0.21 &$-$0.80 &$-$1.92 &$-$3.33 &$-$4.44 &$-$4.72 \\
$-$0.4 &  4.0 &  0.80 &  1.05 &  0.84 &  0.158 &  3.27 &  1.97 &  0.24 &$-$0.83 &$-$2.26 &$-$4.38 &$-$5.58 &$-$6.06 \\
$-$0.4 &  5.0 &  0.82 &  1.07 &  0.84 &  0.167 &  3.43 &  2.07 &  0.35 &$-$0.72 &$-$2.14 &$-$4.22 &$-$5.41 &$-$5.88 \\
$-$0.4 &  6.3 &  0.84 &  1.09 &  0.84 &  0.177 &  3.56 &  2.15 &  0.43 &$-$0.63 &$-$2.05 &$-$4.05 &$-$5.24 &$-$5.69 \\
$-$0.4 &  7.9 &  0.86 &  1.10 &  0.83 &  0.186 &  3.68 &  2.26 &  0.55 &$-$0.51 &$-$1.93 &$-$3.90 &$-$5.09 &$-$5.54 \\
$-$0.4 & 10.0 &  0.88 &  1.12 &  0.83 &  0.197 &  3.74 &  2.32 &  0.63 &$-$0.42 &$-$1.84 &$-$3.77 &$-$4.96 &$-$5.38 \\
$-$0.4 & 11.2 &  0.89 &  1.13 &  0.84 &  0.201 &  3.74 &  2.35 &  0.68 &$-$0.37 &$-$1.79 &$-$3.73 &$-$4.91 &$-$5.34 \\
$-$0.4 & 12.6 &  0.90 &  1.14 &  0.84 &  0.205 &  3.73 &  2.38 &  0.72 &$-$0.32 &$-$1.74 &$-$3.69 &$-$4.87 &$-$5.30 \\
$-$0.4 & 14.1 &  0.91 &  1.16 &  0.84 &  0.212 &  3.72 &  2.40 &  0.76 &$-$0.28 &$-$1.70 &$-$3.64 &$-$4.82 &$-$5.25 \\
$-$0.4 & 15.8 &  0.92 &  1.17 &  0.84 &  0.219 &  3.71 &  2.44 &  0.80 &$-$0.23 &$-$1.65 &$-$3.55 &$-$4.72 &$-$5.13 \\
$-$0.4 & 17.8 &  0.93 &  1.18 &  0.84 &  0.223 &  3.46 &  2.38 &  0.80 &$-$0.22 &$-$1.64 &$-$3.48 &$-$4.65 &$-$5.05 \\
\ppm0.0 &  4.0 &  0.89 &  1.13 &  0.89 &  0.218 &  4.10 &  2.68 &  0.99 &$-$0.04 &$-$1.53 &$-$4.09 &$-$5.29 &$-$5.80 \\
\ppm0.0 &  5.0 &  0.90 &  1.14 &  0.89 &  0.224 &  4.20 &  2.79 &  1.11 &  0.08 &$-$1.41 &$-$3.93 &$-$5.14 &$-$5.63 \\
\ppm0.0 &  6.3 &  0.91 &  1.15 &  0.89 &  0.232 &  4.41 &  2.93 &  1.25 &  0.23 &$-$1.25 &$-$3.79 &$-$4.99 &$-$5.48 \\
\ppm0.0 &  7.9 &  0.93 &  1.18 &  0.89 &  0.244 &  4.57 &  3.02 &  1.34 &  0.33 &$-$1.16 &$-$3.71 &$-$4.92 &$-$5.41 \\
\ppm0.0 & 10.0 &  0.96 &  1.20 &  0.90 &  0.258 &  4.70 &  3.09 &  1.42 &  0.42 &$-$1.05 &$-$3.61 &$-$4.81 &$-$5.30 \\
\ppm0.0 & 11.2 &  0.97 &  1.22 &  0.90 &  0.265 &  4.75 &  3.11 &  1.46 &  0.46 &$-$1.00 &$-$3.59 &$-$4.79 &$-$5.29 \\
\ppm0.0 & 12.6 &  0.99 &  1.24 &  0.91 &  0.273 &  4.80 &  3.14 &  1.49 &  0.50 &$-$0.96 &$-$3.56 &$-$4.76 &$-$5.26 \\
\ppm0.0 & 14.1 &  1.01 &  1.25 &  0.92 &  0.280 &  4.81 &  3.15 &  1.52 &  0.54 &$-$0.92 &$-$3.55 &$-$4.75 &$-$5.24 \\
\ppm0.0 & 15.8 &  1.02 &  1.26 &  0.92 &  0.287 &  4.83 &  3.17 &  1.56 &  0.59 &$-$0.86 &$-$3.53 &$-$4.72 &$-$5.22 \\
\ppm0.0 & 17.8 &  1.03 &  1.28 &  0.92 &  0.294 &  4.83 &  3.20 &  1.60 &  0.64 &$-$0.81 &$-$3.50 &$-$4.70 &$-$5.20 \\
$+$0.2 &  4.0 &  0.94 &  1.17 &  0.92 &  0.252 &  4.44 &  2.99 &  1.32 &  0.33 &$-$1.11 &$-$3.95 &$-$5.16 &$-$5.64 \\
$+$0.2 &  5.0 &  0.94 &  1.17 &  0.91 &  0.256 &  4.53 &  3.13 &  1.47 &  0.47 &$-$0.97 &$-$3.82 &$-$5.03 &$-$5.52 \\
$+$0.2 &  6.3 &  0.97 &  1.20 &  0.92 &  0.269 &  4.78 &  3.25 &  1.57 &  0.57 &$-$0.88 &$-$3.74 &$-$4.94 &$-$5.43 \\
$+$0.2 &  7.9 &  0.99 &  1.22 &  0.93 &  0.282 &  4.95 &  3.33 &  1.66 &  0.68 &$-$0.75 &$-$3.73 &$-$4.95 &$-$5.46 \\
$+$0.2 & 10.0 &  1.02 &  1.25 &  0.92 &  0.296 &  5.10 &  3.42 &  1.76 &  0.78 &$-$0.63 &$-$3.52 &$-$4.72 &$-$5.21 \\
$+$0.2 & 11.2 &  1.03 &  1.26 &  0.93 &  0.305 &  5.15 &  3.42 &  1.77 &  0.81 &$-$0.58 &$-$3.50 &$-$4.71 &$-$5.20 \\
$+$0.2 & 12.6 &  1.05 &  1.28 &  0.94 &  0.314 &  5.19 &  3.43 &  1.80 &  0.85 &$-$0.53 &$-$3.49 &$-$4.70 &$-$5.20 \\
$+$0.2 & 14.1 &  1.06 &  1.29 &  0.94 &  0.323 &  5.23 &  3.45 &  1.83 &  0.89 &$-$0.47 &$-$3.48 &$-$4.69 &$-$5.19 \\
$+$0.2 & 15.8 &  1.08 &  1.31 &  0.95 &  0.331 &  5.25 &  3.48 &  1.87 &  0.94 &$-$0.43 &$-$3.47 &$-$4.68 &$-$5.18 \\
$+$0.2 & 17.8 &  1.09 &  1.33 &  0.95 &  0.338 &  5.29 &  3.52 &  1.92 &  0.99 &$-$0.38 &$-$3.46 &$-$4.67 &$-$5.17 \\
\hline
\end{tabular}\end{minipage}
\end{table*}

\subsection{SBF Predictions}

The luminosity-weighted mean luminosity \lbar\ of a stellar 
population in a
photometric band $\lambda$ is given by the ratio of the second 
and first moments of the stellar luminosity function $n(L_\lambda)$
in that band:
\begin{equation}
\overline{L}_\lambda \,=\, 
{\int{n(L_\lambda)\,L_\lambda^2 \, dL_\lambda}\over
{\int{n(L_\lambda)\,L_\lambda \, dL_\lambda}}}
\,=\, {\langle L_\lambda^2 \rangle \over \langle L_\lambda \rangle}
\;.\label{eq:lbar}
\end{equation}
This luminosity is then converted to the magnitude \Mbar.
Because of the luminosity weighting, \Mbar\ is most
sensitive to the brightest stars in a population.
See Liu \etal\ (2000) for a quantitative discussion of the
contributions of stars at different evolutionary phases to 
the SBF signal.

\begin{figure*}
\epsfig{figure=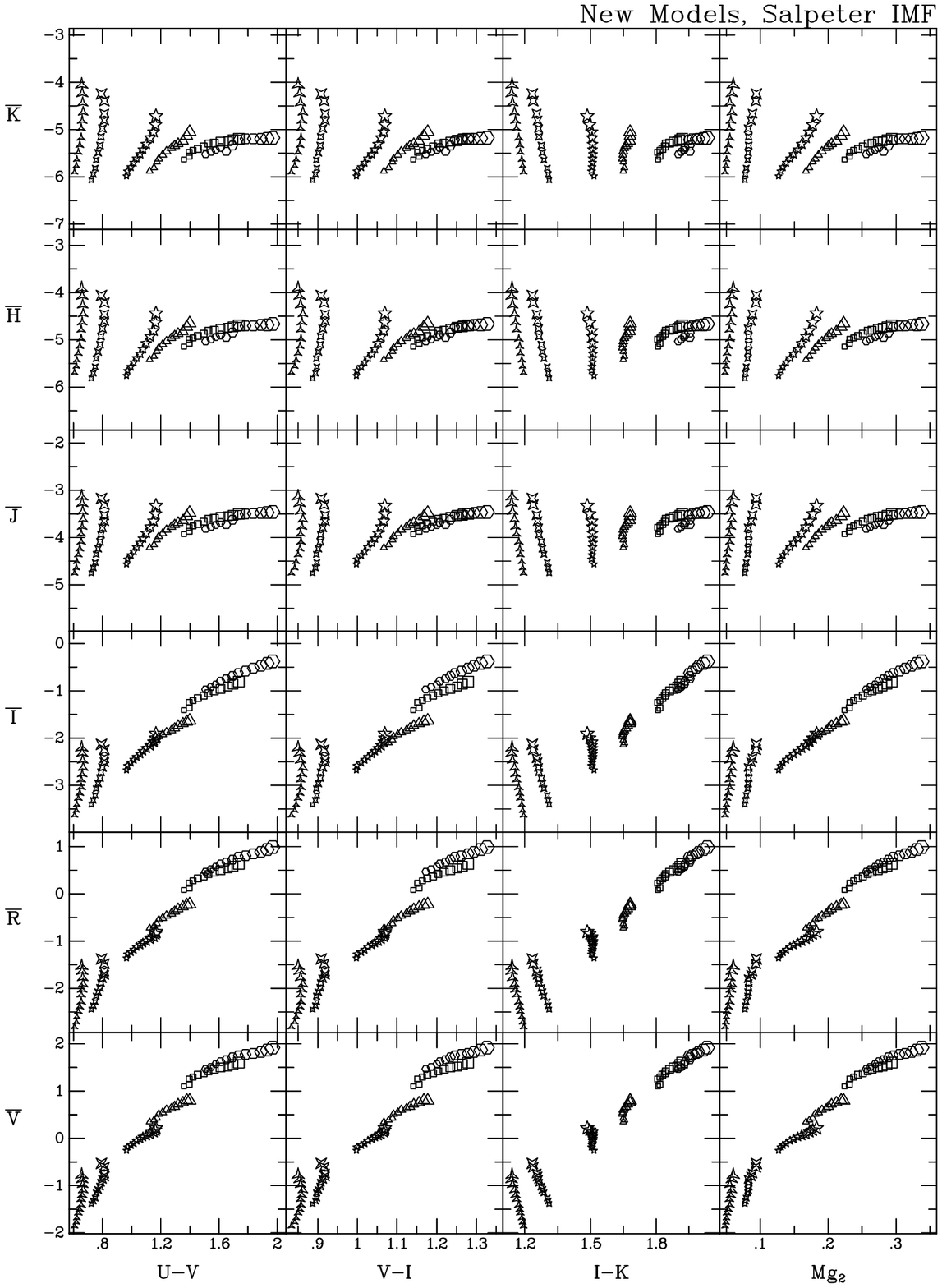, height=1.21\textwidth, width=0.88\textwidth}
\medskip
\caption{Various optical and near-IR SBF magnitudes for the Vazdekis (2000)
models, using the new Padua isochrones (Girardi \etal\ 2000) and a Salpeter
IMF, are plotted against three different integrated colours and the Mg$_2$
index.  The vertical scale is the same for all the plots.  The symbols are
based on the convention used by Worthey (1993a, 1994) and are coded
according to metallicity [Fe/H] as follows: $-1.7$ (three-pointed stars),
$-1.3$ (four-pointed stars), $-0.7$ (five-pointed stars), $-0.4$
(triangles), $0.0$ (squares), $+0.2$ (hexagons).  The symbol size is
according to age (bigger symbols for greater ages).  All isochrone ages of
5.0\,Gyr or more (increasing in steps of 12\% to 17.8\,Gyr) are shown.
}\label{fig:newcals}
\end{figure*}

\begin{figure*}
\epsfig{figure=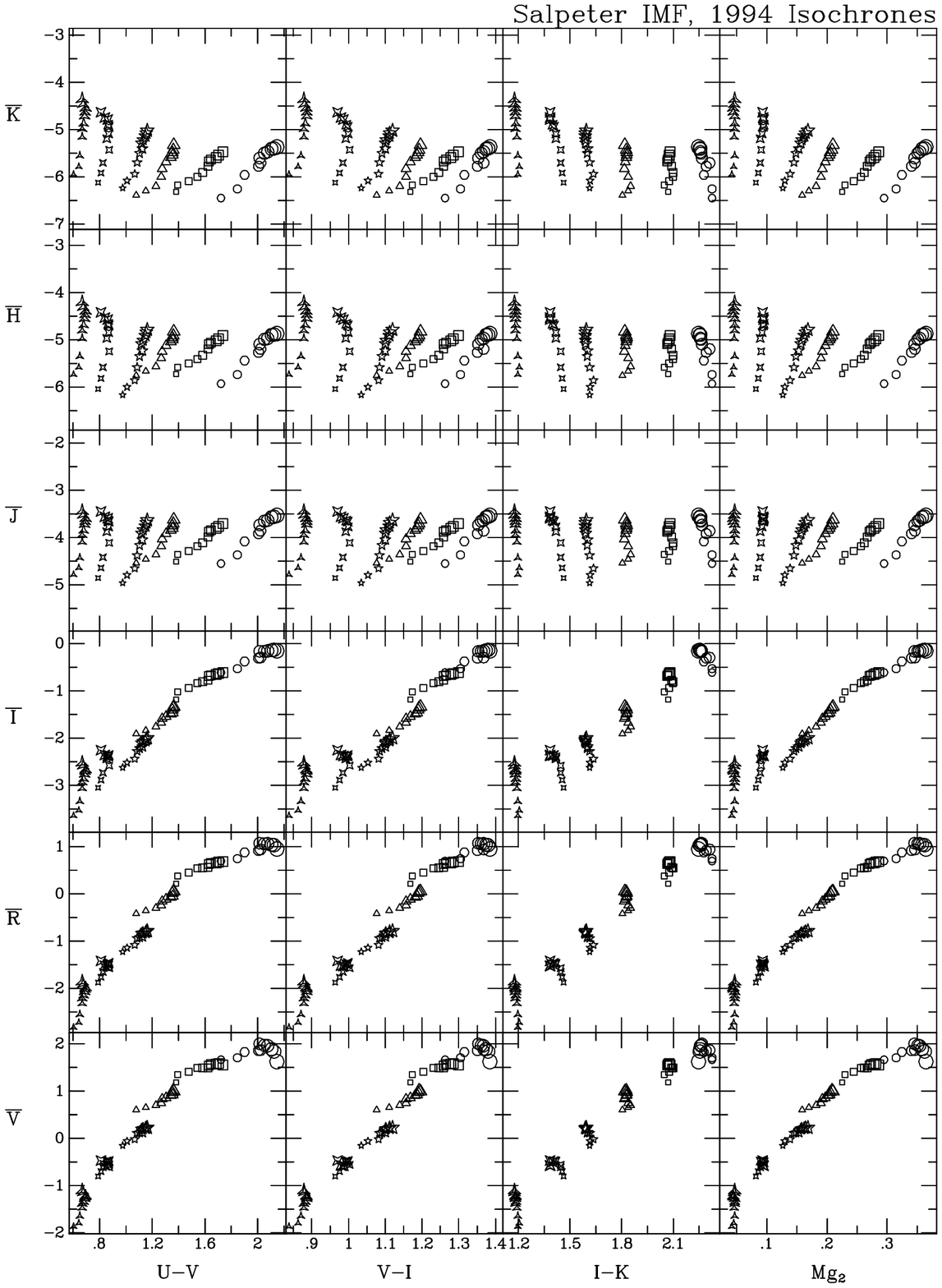, height=1.21\textwidth, width=0.88\textwidth}
\medskip
\caption{ Similar to Figure~\ref{fig:newcals}, except that the new Padua
isochrones have been replaced with the `old' ones (Bertelli \etal\ 1994) in
the models.  The highest metallicity model here is $\hbox{[Fe/H]}=+$0.4
dex, which we represent with circles; otherwise the metallicities and
symbols are the same as in Figure~\ref{fig:newcals}.  The vertical limits
are identical to those in Figure~\ref{fig:newcals} for ease of comparison.
Again, all isochrone ages of 5.0\,Gyr or more are shown, but the age steps
are irregular and the maximum age is 17.4\,Gyr.  These plots demonstrate
that simply replacing the isochrone set with another, similar, set results
in significant differences.  }\label{fig:oldcals}
\end{figure*}

\subsubsection{New Models Results}

We have calculated SBF magnitudes for the models
described in the previous section using a range of IMFs.
As an example,
Table~\ref{tab:sbf_un1.3} lists some results for models with a unimodal IMF
of slope $\mu = 1.3$ (Salpeter).  The table lists the metallicity (dex)
and age (Gyr) of each model, several broadband colours, the Mg$_2$
absorption index (mag), and the SBF magnitudes in the $UBVRIJHK$ bandpasses.  
Results are given for ten different ages and all six metallicities.  
The chief limitation of these models is their restricted metallicity
range. In particular, a large fraction of Galactic globular clusters have
metallicities $\hbox{[Fe/H]} < -1.7$, and thus cannot be compared with
these models. The choice of ages comes from Girardi \etal\ (2000),
except we omit models younger than 4\,Gyr and age steps of less than
1\,Gyr.  For ages less than $\sim\,$5\,Gyr, the highly uncertain
thermally pulsating AGB stage in the model isochrones becomes more
luminous than the tip of the red giant branch (RGB), and thus the SBF
magnitudes become quite uncertain.  Worthey (1993a) likewise noted this
point and reported results for just the 5\,Gyr and older models.  
The complete model output is available on request to the 
authors\footnote{or see http://star-www.dur.ac.uk/$\sim$vazdekis/}.

Figure~\ref{fig:newcals} plots the $VRIJHK$ SBF magnitudes for the new
model SSPs using a Salpeter IMF against three integrated colours and \mgii.
As expected, the optical SBF magnitudes become fainter with increasing
metallicity and redder integrated colour.  The effects of metallicity
and age on the optical \Mbar-colour relations are less
degenerate than they were in the Worthey models, and thus these
SSP plots show more scatter than those ones.  The \Mi-\mgii\ relation
does however define a very narrow locus at intermediate
and high metallicity due to metallicity-age degeneracy.
At the two lowest model metallicities, appropriate for
Galactic globular clusters, the degeneracy is much less:
\Mbar\ becomes mainly sensitive to age while colour
and \mgii\ depend more on metallicity.

We do not plot the results for $\overline M_U$ and $\overline M_B$ as these
are too faint to be useful in extragalactic studies and there are almost no
data available.  However, the $\overline M_B$ plots look very similar to
the $\vbar$ ones, except shifted fainter by 1.65\,mag. The $\overline M_U$
predictions are down by another $\sim$1.3\,mag and are significantly
different, with more age-metallicity degeneracy at low metallicities and
less at high metallicities.

In the near-IR bands, \Mbar\ shows much less total variation, especially
for the higher metallicity models, which indicate that \jbar, \hbar, and \kbar\
should all be excellent distance indicators for metal-rich giant ellipticals.
However, at lower model metallicities, age and metallicity are less degenerate.
For instance, at $\hbox{[Fe/H]}<0$, the
\MbJ-$(I{-}K)$ plot completely breaks the degeneracy, with \MbJ\ being a
sensitive age indicator and $(I{-}K)$ varying almost exclusively with
metallicity.  This could be useful for disentangling these parameters in
stellar population studies, but may add uncertainty to the distance
measurements of bluer ellipticals.

Of course, the accuracy of the distance indicator depends
on the mix of the simple stellar populations within actual stellar
systems.  We discuss this problem more in Sec.\,\ref{sec:theorydists}.
Here we examine some of the variables and uncertainties with the SSP 
predictions themselves.

\subsubsection{Results for Previous Isochrones}

Figure~\ref{fig:oldcals} displays results as in
Figure~\ref{fig:newcals} but for models which simply replace the new
Girardi \etal\ (2000) isochrones with the older set of Padua isochrones
(Bertelli \etal\ 1994, augmented by Pols \etal\ 1995), which at fixed
age and metallicity have a cooler giant branch and lower main sequence
but a hotter turnoff. The `synthetic' AGB phases are also different.
The differences in the resulting SBF predictions are significant.
The mean SBF magnitude gets 0.1--0.2\,mag fainter in the
optical and $\sim\,$0.5\,mag brighter in the near-IR.  The
age-metallicity degeneracy in the optical $\Mbar$--colour relations is
greater for these models, and the scatter is consequently decreased.  In
the near-IR, the breaking of the degeneracy persists to the highest
metallicities and reddest integrated colours.  Thus, these models would
indicate that SBF should show large scatter in the near-IR even for
giant ellipticals.

We tested to see how much of the discrepancy between the two sets of model
predictions is due to differences in the AGB prescriptions of Girardi
\etal\ (2000) and Bertelli \etal\ (1994).  We calculated solar metallicity
models of age 10 Gyr but omitting stars in the AGB phase.  The differences
between the optical SBF predictions of the two sets of models actually
increased by a few hundredths of a magnitude, but the differences between
the near-IR SBF predictions decreased by $\sim70$\%.  Thus, the AGB
prescription accounts for most (though not all) of the discrepancy
between Figures~\ref{fig:newcals} and~\ref{fig:oldcals} in the near-IR,
but not in the optical.  The discrepancy in the optical, and the rest of
the near-IR discrepancy, must result primarily from the RGB track
calculations.

It is sobering to note that the fine tuning of the
stellar physics and the revision of the AGB prescription
by Girardi \etal\ has resulted in a markedly
different set of SBF predictions.  SBF is sensitive to the brightest
stars in a stellar population, and in an old system these are
evolved stars, whose properties tend to be the most uncertain.
More optimistically, one could say that the sensitivity of SBF to these 
stars provides the opportunity for new constraints on the stellar evolution,
particularly at high metallicity where there are no
Galactic globular clusters.

For the remainder of this paper we mainly restrict the discussion to the
models which use the new isochrones of Girardi \etal\ (2000) because of the
improvements in the stellar physics.  However, we note where there
would be major differences in the results and interpretations if models
based on the previous set of Padua isochrones had been used.

\subsubsection{Effect of the Empirical Transformations}

\begin{figure}
\epsfig{figure=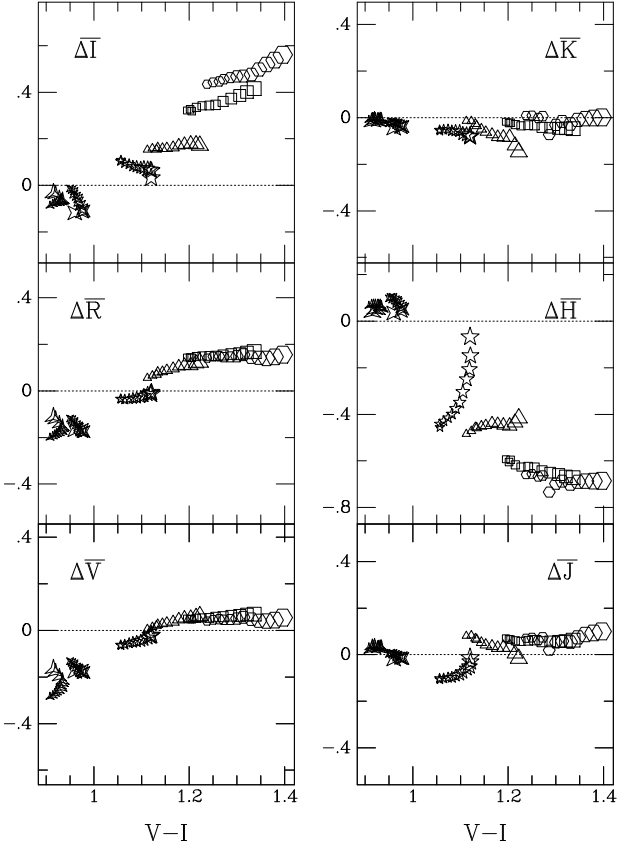, height=0.63\textwidth, width=0.46\textwidth}
\medskip
\caption{Differences between our (Vazdekis 2000) model SBF predictions
and the SBF magnitudes calculated using the integrated magnitudes and
colours supplied with the Girardi \etal\ (2000) isochrones. 
Symbols are as in Figure~\ref{fig:newcals}.  Since the isochrones are the
same, the differences are due to the different set of transformations from
the theoretical to the observational plane.  Positive (negative) differences 
indicate that the Vazdekis SBF magnitudes are fainter (brighter).
}\label{fig:dmbar}
\end{figure}

We remarked above that the empirical approach to transforming between the
theoretical and the observational planes is one of the major strengths of
the models used here.  In order to test the importance of these empirical
transformations, we have computed SBF magnitudes using the mainly
theoretically based photometric fluxes supplied with the Padua isochrones
(Girardi \etal\ 2000).  Figure~\ref{fig:dmbar} shows the differences
between the SBF magnitudes predicted by the Vazdekis models and by the
Girardi \etal\ isochrones with the supplied fluxes.
While the \vbar, \jbar, and \kbar\ predictions differ at the 0.1\,mag
level, the \vbib\ fluctuation colours differ at high metallicity by as
much as 0.5\,mag, with the theoretical predictions being redder (larger).
The biggest differences occur for two of the most widely used SBF
bandpasses: $(\ibar-\hbar)$ changes by up to 1.2\,mag at high metallicity
between the two sets of predictions.

The near-IR spectra of metal-rich giant stars are full 
of molecular absorption features; the slope of the theoretical
energy distribution will be sensitive to how well these are modeled.
Thus, it is not surprising that the empirical transformations
give quite different SBF results in the near-IR than the mainly 
theoretical transformations, although it is unclear why the $J$ 
and $K$ predictions agree so well.
Lejeune \etal\ (1997) compared their empirical stellar colours
to theoretical colours from model atmospheres. Their
plots show that the most significant systematic problems occur
in the $B$, $I$, and $H$ bands.
However, the transformations supplied with the Padua isochrones
differ from those considered by Lejeune et al.\ (1997).

In any case,
Figure~\ref{fig:dmbar} vividly illustrates the importance of the empirical
transformations for this study.  Given the size of the differences,
one hopes that the samples on which the empirical transformations are
based will continue to grow in the number and variety of stars used. 
Improvements in the theoretical stellar atmospheric fluxes are also needed.

\subsubsection{The IMF}

\begin{figure*}
\epsfig{figure=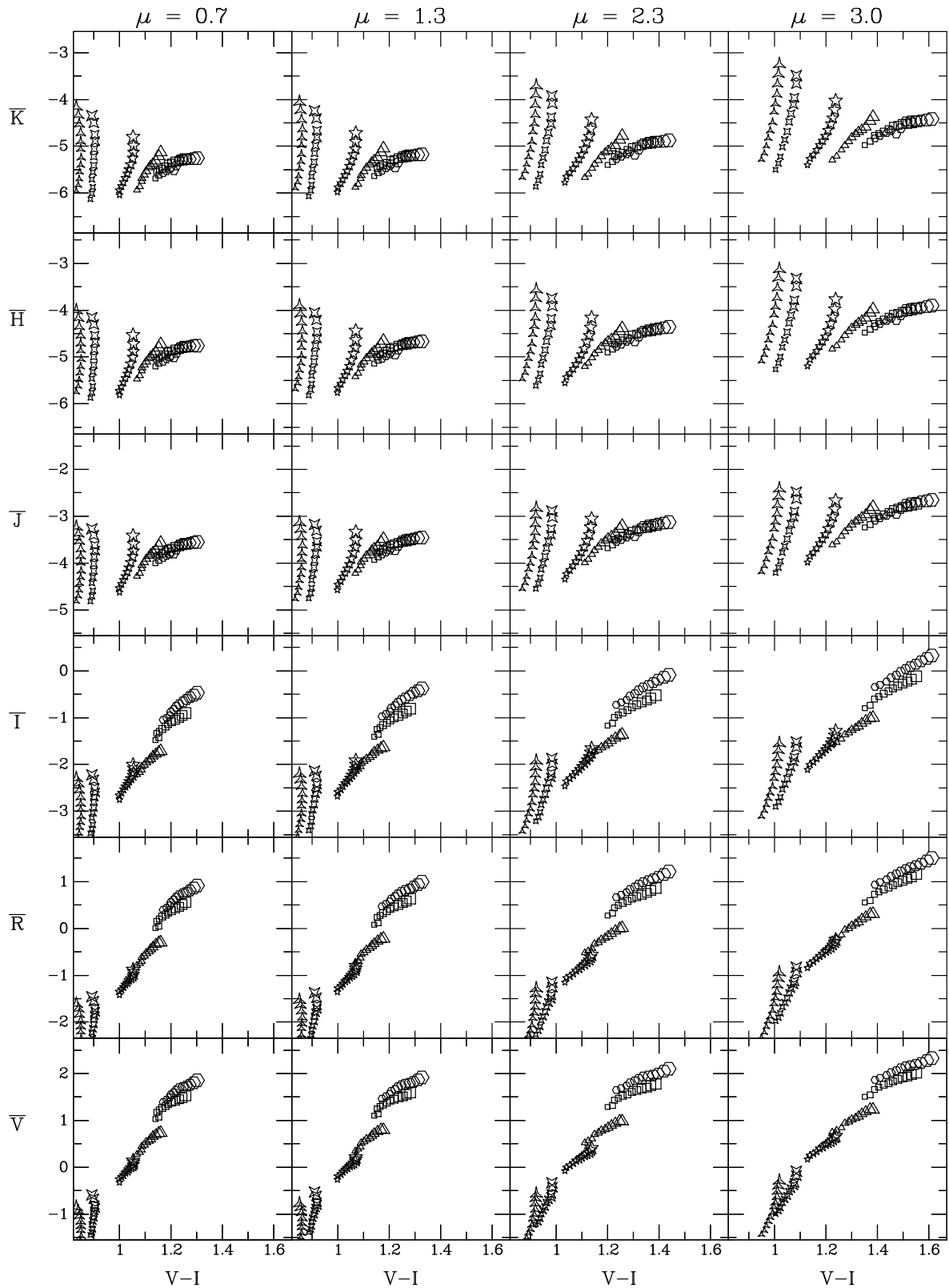, height=1.20\textwidth, width=0.88\textwidth}
\medskip
\caption{Model SBF predictions are plotted against \vi\ for 4 different
unimodal IMFs with logarithmic slopes of 0.7, 1.3 (Salpeter), 2.3, and 3.0.
Symbols as in Figure~\ref{fig:newcals}.
The second column of plots from the left is identical to the second
column in Figure~\ref{fig:newcals} except that the limits are changed for
consistency with the other IMF columns in this figure.
}\label{fig:multiimf}
\end{figure*}

Figure~\ref{fig:multiimf} demonstrates how the \Mbar-\vi\ relations for
the new models depend on the slope of the (unimodal) IMF.  The colours are
redder and the fluctuations fainter with the steeper (more dwarf dominated)
IMFs.  Worthey (1993a) concluded that IMF effects on the optical
relations are nearly degenerate with those of metallicity and age.
For the more moderate unimodal IMFs ($\mu\approx1$--2), \Mbar\
shows relatively little change in the optical at a given colour, thus
corroborating Worthey's basic conclusion.  However, in the near-IR
the mean and scatter in \Mbar\ at a given colour changes more
significantly.  

Vazdekis \etal\ (1997) explored a range of both unimodal and bimodal IMFs
in detailed comparisons of the SSP model colours and spectral indices with
data for several elliptical galaxies.  For the unimodal case, they obtained
the best fits with a Salpeter-like slope.  However, the fits using the
bimodal IMF with $\mu\approx2.3$ at intermediate to high masses were
better.  (This result is independent of their conclusions regarding a
variable IMF in the full evolutionary models.)  Since the slope at the
low-mass end was fixed, the bimodality does not introduce an extra degree
of freedom.  This type of IMF, with a fairly steep high-mass slope and flat
at low masses, is quite similar to a Miller-Scalo IMF and closely resembles
the observed mass function of the Pleiades and some other open clusters.
[Scalo (1998) gives a comprehensive review of the observations; see also
recent reviews by Meyer \etal\ (2000) and Paresce \& De\,Marchi (2000).]

\begin{figure*}
\epsfig{figure=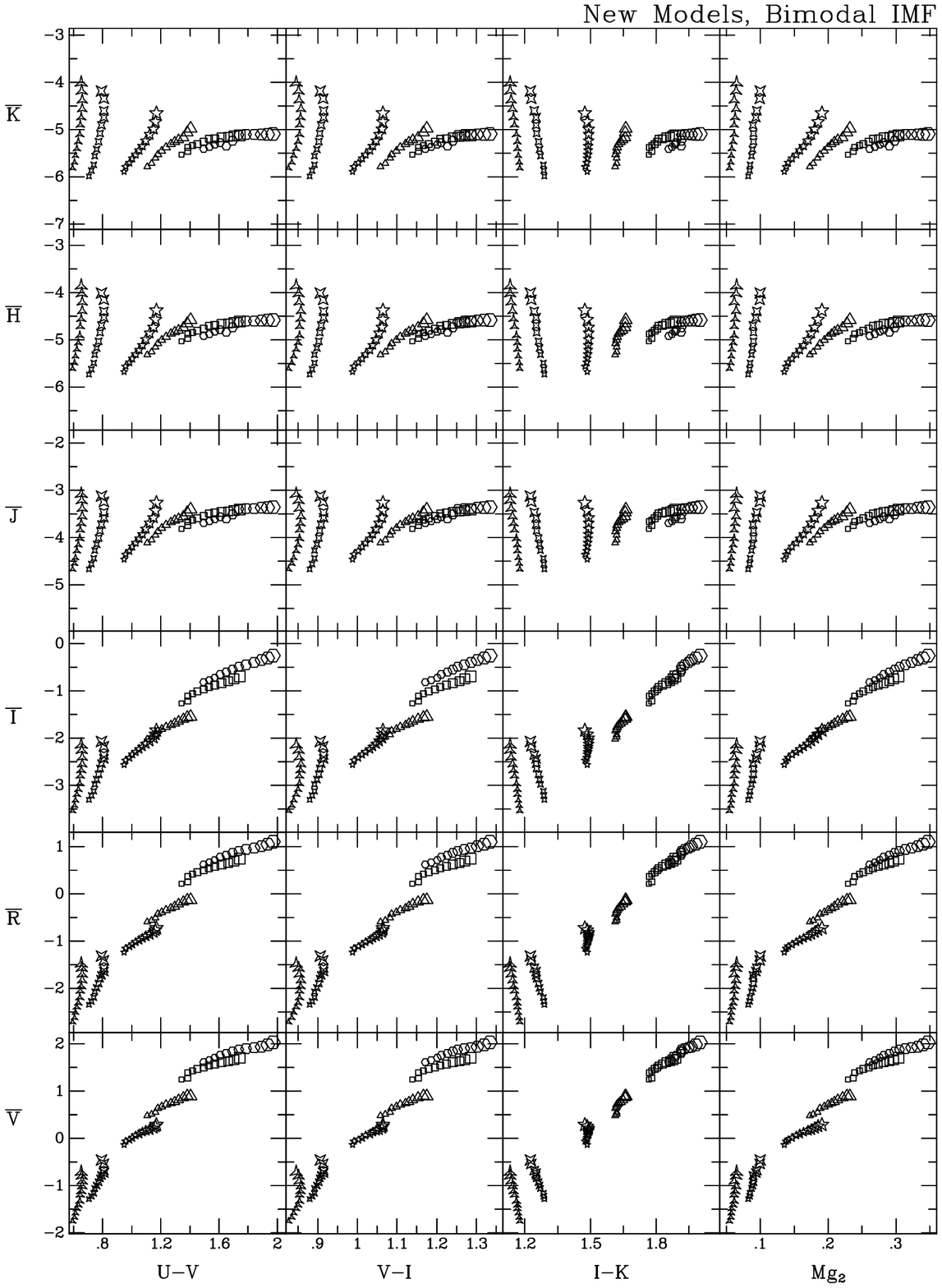, height=1.21\textwidth, width=0.88\textwidth}
\medskip
\caption{ Similar to Figure~\ref{fig:newcals}, except that we use the
bimodal IMF from Vazdekis \etal\ (1996), which is flat (in the logarithmic
slope) below 0.4\,\msun\ and has a steep slope of $\mu=2.3$ above
0.6\,$M_\odot$, with a smooth transition in between.  This approximates a
Miller-Scalo IMF, and the resulting predictions are close to the unimodal
$\mu=1.3$ case illustrated in Figure~\ref{fig:newcals}.  The limits are
identical to those in Figure~\ref{fig:newcals} for ease of comparison.
}\label{fig:bimodcal}
\end{figure*}

Figure~\ref{fig:bimodcal} shows the \Mbar\ predictions of the new models
for this type of IMF.  They are remarkably similar to the results shown
in Figure~\ref{fig:newcals} for the Salpeter IMF, with the main difference
being a shift in \Mbar\ fainter by $\sim\,$0.05\,mag.
The reason is that the decrease in the number of giants
due to the steeper high-mass slope is mostly compensated by the
decrease in the lower main sequence luminosity due to the shallower
low-mass slope.  Put another way, a Salpeter IMF reasonably
approximates a Miller-Scalo IMF.

Thus, the precise form of the IMF appears to be of secondary importance.
The different IMFs give similar \Mbar-colour relations, even though the
model age and metallicity at a fixed point in the $(\Mbar, [V{-}I])$ plane
varies.  For the sake of convention, we concentrate throughout
most of the rest of this paper on the results for a Salpeter IMF,
commenting where necessary on the effect of different IMFs.

\section{Comparing Models with Data}
\label{sec:disc}

\subsection{Simple Stellar Populations: Globular Clusters}

\begin{figure}
\epsfig{figure=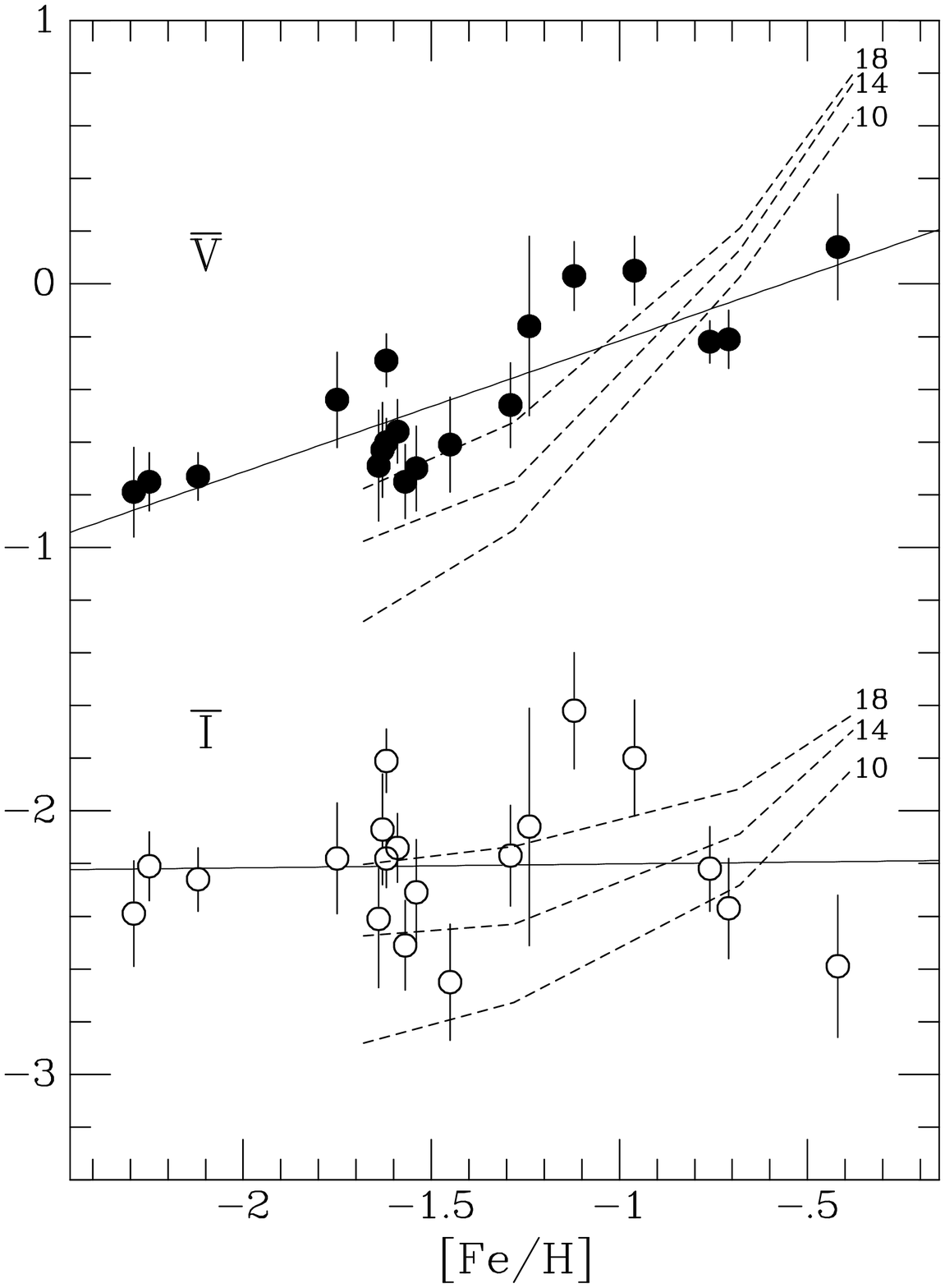, height=0.63\textwidth, width=0.46\textwidth}
\medskip
\caption{Comparison of the \MbV\ and \MbI\ measurements for Galactic
globular clusters by Ajhar \& Tonry (1994) with the new model
(Vazdekis 2000; Salpeter IMF) predictions (dashed lines).
The models are for ages of 10.0, 14.1, and 17.8 Gyr (labelled),
with the younger ages giving brighter SBF magnitudes.
The [Fe/H] values for the globulars are taken from the Harris
(1996) electronic catalogue, version dated 22~June 1999, and we have
adopted the horizontal branch calibration used in that catalogue to set
the distances.  The solid lines are simple linear fits to the data.
}\label{fig:globcal}
\end{figure}

The least ambiguous way to test SSP models is to compare their
predictions to the observed properties of globular clusters.
Figure~\ref{fig:globcal} compares the Ajhar \& Tonry (1994) results for
\vbar\ and \ibar\ in 19 Galactic globular clusters with the SBF
predictions of our new models.

The globular cluster distances were calculated from 
the `compromise' relation
between the horizontal branch magnitude and metallicity
advocated in the current (June 1999)
version of the Harris (1996) electronic catalogue:
\begin{equation}
 M_V({\rm HB}) \,=\, 0.15 \hbox{[Fe/H]} + 0.80\,.
\label{eq:RRlyra}
\end{equation}
The slope here is the same as used in the analysis by Ajhar \& Tonry,
who adopted the calibration of Carney, Storm, \& Jones (1992) based on
statistical parallax and Baade-Wesselink measurements of field RR Lyrae
stars, but the zero~point is brighter by 0.21\,mag.  However, this
zero~point is 0.1--0.2\,mag {\it fainter} than those favoured by 
recent analyses of Hipparcos subdwarf parallaxes (Gratton \etal\ 1997;
Carretta, \etal\ 2000), but it accords well with the calibration derived
when the RR Lyrae luminosity is set by requiring consistency with the
Cepheid distance to the Large Magellanic Cloud (e.g. Walker 1992).
Carretta \etal\ (2000) make note of this last point and so adjust their
final calibration fainter by 0.1\,mag, concluding that the distances
derived from the subdwarf fitting are about $1\sigma$ too large.

All of the above methods for deriving the HB/RR~Lyra calibration, as
well as HB measurements in M31 globulars (Ajhar \etal\ 1996; Fusi Pecci
\etal\ 1996), agree fairly well on a slope of 0.15$\pm$0.05, but values
of 0.30 or more have also been proposed (e.g., Sandage 1993).  
Using a distance calibration with a steeper slope makes
the derived slope of \Mbar\ vs [Fe/H] steeper, and thus more similar to
the model predictions.  However, as there are more direct and less
model-dependent means for deriving the slope of the RR~Lyra calibration,
we do not allow ourselves this added freedom.

With the distance calibration from Eq.\,(\ref{eq:RRlyra}),
the Ajhar \& Tonry (1994) globular cluster data give
\begin{eqnarray}
\overline M_V &=& -0.46 \;+\; 0.44\,(\hbox{[Fe/H]} + 1.5) \nonumber \\
\overline M_I &=& -2.21 \,,
\label{eq:globobs}
\end{eqnarray}
with rms scatters of 0.16 and 0.26 mag in $V$ and $I$, respectively.  There
is no significant slope in the $I$~band.  The models agree well with the
$I$-band data in zero~point and are marginally too bright
(given the $\sim\,$0.2\,mag uncertainty in the observational zero point)
in the $V$~band, where the metal-poor globulars would require
greater model ages. Both data and models exhibit a much steeper
dependence for \Mv\ on metallicity. 
The models predict an increasingly shallow slope at increasingly
lower metallicities in both bands.

We find the agreement between the globular cluster SBF data and the new
models encouraging.  The agreement is further improved if the
more metal-rich globulars are also younger, or, stated another way,
this is what the present comparison predicts.  The baseline of this
comparison needs to be increased with more SBF measurements of higher
metallicity globulars (unfortunately these are generally at large distances
and low Galactic latitudes, with severe extinction and disk/bulge star
contamination), as well as by extending the model predictions to 
lower metallicities.

\begin{figure}
\epsfig{figure=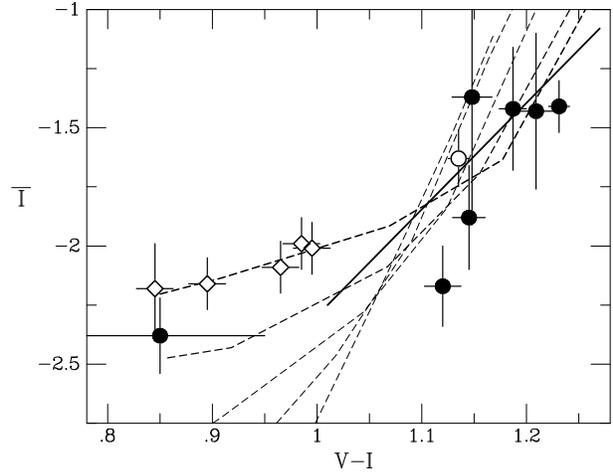, height=0.35\textwidth, width=0.45\textwidth}
\medskip
\caption{Comparison of \MbI\ model predictions with measurements for
galaxies with Cepheid distances (data from SBF-III).  Filled circles are
for spirals with bulge \mi\ measurements and for the dwarf irregular
NGC\,5253, which has $\vi{\,=\,}0.85$.  Open symbols are for the compact
elliptical M32 (circle) and various locations
(10--130 arcsec) in the dwarf elliptical
NGC\,205 (diamonds), both assumed to be at the same Cepheid distance as M31.
(The M32 data point represents a mean within a radius of about 2.5~arcmin.)
The dashed lines connect models of different metallicity but fixed
age; the ages represented are 4.0, 7.1, 10.0, 14.1, and 17.8 Gyr, with
heavier lines used for greater ages.  The solid line shows the
empirically calibrated relation based on SBF measurements to ellipticals
in groups with Cepheids distances.  }\label{fig:cephcal}
\end{figure}

\subsection{SBF in Galaxies with Cepheid Distances}
\label{ssec:sbfcephs} 

Because the method works best for ellipticals, SBF distances are difficult
to calibrate against Cepheids.  We are forced to attempting the analysis in
irregulars and spiral bulges or else associating early- and late-type
galaxies via uncertain group membership.  Both options are fraught with
systematic problems (see Appendix~B of SBF-II) but are currently the only
means for comparing our models against the Cepheid distance scale.
Figure~\ref{fig:cephcal} shows the comparison for the $I$~band, the only
band with direct SBF measurements in Cepheid-bearing galaxies other than
M31.  Two of the galaxies in Figure~\ref{fig:cephcal} have \MbI\ values too
bright by more than 1$\,\sigma$ as compared to the models. Otherwise the
agreement appears reasonably good.

The flattening of the \MbI-\vi\ relation for galaxies with $\vi<1.0$ was
noted as a puzzle by SBF-I, and these galaxies were excluded from the
calibration, but our new SSP models predict this effect.  The apparent
18-Gyr age for NGC\,205 is surely too high, but this may be misleading due
to composite population effects.  The combination of old very metal-poor
stars with a young high-metallicity component could yield observed position
for NGC\,205 in Figure~\ref{fig:cephcal}.  In addition, the true distance of
this galaxy may well be greater than the assumed M31 Cepheid distance.
Observations of both RR~Lyra stars (Saha, Hoessel, \& Krist 1992) and
post-AGB stars (Fullton \& Bond 1997) indicate that NGC\,205 is about
0.3~mag more distant than M31; if so, its inferred age would decrease by
5\,Gyr.

M32 is the only `typical SBF galaxy' in Figure~\ref{fig:cephcal},
and it appears consistent with
the models if placed at the same distance as M31.  However, recent
spectroscopic population synthesis studies of this galaxy find 
best-fitting models with solar metallicity, or slightly below, and an age
typically 4--6\,Gyr (Jones \& Worthey 1995; Vazdekis \& Arimoto 1999; 
del\,Burgo \etal\ 2000; Trager \etal\ 2000a).  Our models with this age
and metallicity (see Table~\ref{tab:sbf_un1.3}) match the observed 
$\vi=1.13$ colour of this galaxy but give \MbI\ too faint by 0.1--0.2\,mag.
Of course, the spectroscopic analyses used much smaller apertures
than the SBF analysis.  Rose \etal\ (2000, in preparation) find that
M32 has a radial age/metallicity gradient, even though the optical colours
stay nearly constant; thus, a direct comparison to the spectroscopic
measurements is not possible.

Despite the potential problems, the group-based empirical relation shown
in the figure has encouraging overlap with the models.
Section~\ref{sec:theorydists} returns in greater detail to the issues of
composite populations and differences between the empirical and model
SBF calibrations in various bandpasses.

\subsection{Fluctuation Colour versus Integrated Colour}

\begin{figure}
\epsfig{figure=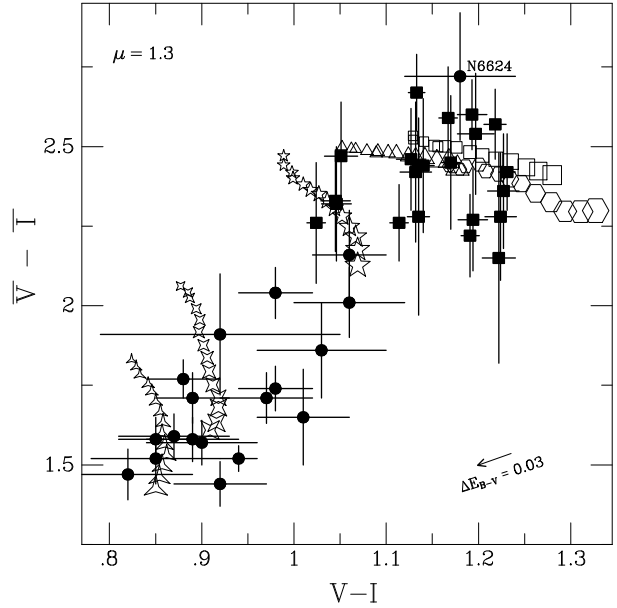, height=0.45\textwidth, width=0.45\textwidth}
\medskip
\caption{The \vbib\ fluctuation colour is plotted against integrated
\vi\ for the new models with ages $\ge4.0\,$Gyr
(symbols coded as in Figure~\ref{fig:newcals}),
the Ajhar \& Tonry (1994) globular clusters (filled circles), and
the data available for 21 galaxies in the Local Group, Leo, Virgo, and
Fornax (filled squares). The models here use a Salpeter ($\mu=1.3$) IMF.
The globular cluster NGC\,6624 (marked) with $\hbox{[Fe/H]}=-0.4$
has the highest metallicity in the Ajhar \& Tonry sample.  
The arrow shows the effect of increasing the extinction estimate
by 0.03\,mag.
}\label{fig:vbib_vi}
\end{figure}

Fluctuation colours are different from integrated colours.  For
instance, Figure~\ref{fig:vbib_vi} plots \vbib\ against \vi\ for 
globular clusters (Ajhar \& Tonry 1994), galaxies (Tonry \etal\ 1990;
this work; SBF-IV), and our new models.  Although there is an overall
correlation between \vbib\ and \vi, these quantities have opposite
age dependences in our models.  As age increases at a fixed metallicity,
the model \vbib\ colour gets bluer while \vi\ gets redder.  This is a fairly
common feature of fluctuation colours in our models.  For instance,
even though $(V{-}K)$ is nearly independent of age at lower metallicities,
\vbkb\ gets nearly a magnitude bluer as age is increased from 5 to 17 Gyr.

\begin{figure}
\epsfig{figure=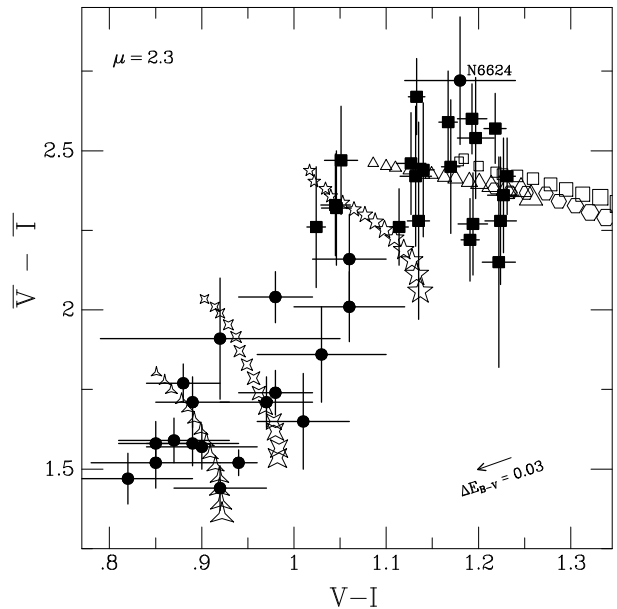, height=0.45\textwidth, width=0.45\textwidth}
\medskip
\caption{Same as Figure~\ref{fig:vbib_vi}, but for a steeper IMF of
unimodal logarithmic slope $\mu=2.3$; the model points have moved
significantly redder in \vi\ and slightly bluer in \vbib, mimicking the
effects of age.  Although the steep IMF makes the match between models and
data better here, other problems arise as a consequence (see text).
}\label{fig:vbib_vi2.3}
\end{figure}

The galaxies in Figure~\ref{fig:vbib_vi} lie close to the higher metallicity
models, where it becomes more difficult to separate age and metallicity
effects.  A careful study of radial population gradients within individual
ellipticals using this diagram would be interesting.  Gradients in
single-band fluctuation magnitudes are well known (Tonry 1991; Sodemann \&
Thomsen 1995, 1996), and simply follow the integrated colour gradients,
which tend to be small for giant ellipticals.  A significant gradient in the
\vbib\ fluctuation colour would be much more difficult to detect because of
the small intrinsic spread and the large uncertainties.  However, it might
reveal whether the radial population gradient is simply a metallicity
gradient or if age gradients are important as well.  With the expected
high-resolution, wide-field imaging capability of future {\it Hubble Space
Telescope} ({\it HST}) instruments, this kind of study should soon be more
feasible.

The globular cluster data in Figure~\ref{fig:vbib_vi} mainly lie
among the $\hbox{[Fe/H]}\le-0.4$, $\hbox{age}\gta10$ Gyr models,
but with an apparently significant age spread.
For instance, near $\vi=0.9$ the data show a significant range
in \vbib, and although zero-point problems are likely,
the trend of the models clearly suggests age variations.
We note that the globular cluster NGC\,6624, which had the
highest metallicity in the Ajhar \& Tonry sample, falls in a
region of the diagram consistent with its photometrically estimated
metallicity of $\hbox{[Fe/H]} = -0.42$.  This value is higher than the
spectroscopic determinations by 0.3\,dex (Vazdekis 1999) to as much as
0.7\,dex (Origlia \etal\ 1997).  Although the methods for determining
metallicity give discrepant results for this globular, it seems reasonable
that SBF measurements would be most consistent with the photometric value.

Figure~\ref{fig:vbib_vi} also reveals a possible shortcoming of the
models: they do not get blue enough in \vbib\ to match several of the
$\vi>0.9$ globulars (this was also apparent in
Figure~\ref{fig:globcal}).  On the other hand, this discrepancy could
arise from underestimating the extinction towards these globulars, as
illustrated by the reddening vector in the figure.  Presumably the
models would reach these data points for ages $\gta20\,$Gyr, but this is
unrealistically large.  The problem is reminiscent of the
difficulty encountered in trying to match the H$\gamma$ absorption
indices of globulars (Vazdekis \& Arimoto 1999; Gibson \etal\ 1999).

Figure~\ref{fig:vbib_vi2.3} shows that the symptoms of this
problem can be removed with a steeper (unimodal) IMF of slope $\mu=2.3$,
which for this plot has effects similar to increasing the age.  (The
Miller-Scalo-like bimodal IMF of high-mass slope 2.3 basically
reproduces the Salpeter IMF results in Figure~\ref{fig:vbib_vi}).
However, such a steep IMF is inconsistent with observations of the
stellar mass function in globulars (Paresce \& De\,Marchi 2000), and
makes \vi\ too red at a given metallicity.  It also implies ages
$\lta7\,$Gyr for some of the globulars, which is inconsistent
with ages based on the colour-magnitude diagram.  Thus, we
continue to prefer the models based on the Salpeter IMF,
although they appear to overestimate the ages somewhat.

\subsection{Fluctuation Colours and Composite Populations}

\begin{figure}
\epsfig{figure=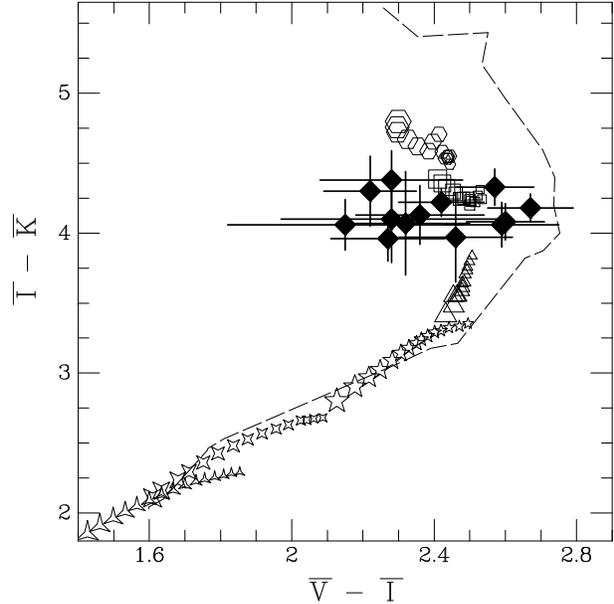, height=0.45\textwidth, width=0.45\textwidth}
\medskip
\caption{The \ibkb\ fluctuation colour is plotted against \vbib\ 
for the new models (symbols as in Figure~\ref{fig:newcals}) and
data for 12 ellipticals and the bulge of M31.
The $V$ data are from Tonry \etal\ (1990) and the
present work; the $I$ data are from SBF-III;
the $K$ data are from Luppino \& Tonry (1993) (two galaxies), 
Jensen \etal\ (1998) (ten galaxies), and Pahre \& Mould (1994) (one galaxy).
The locus of the single-burst old stellar population models of Worthey (1994)
is indicated by a dashed line.
}\label{fig:newsbfcol}
\end{figure}

Blakeslee \etal\ (1999a) showed a comparison of the observed
\vbib\ vs \ibkb\ for 10 galaxies with the SSP model predictions
from Worthey (1993a, 1994).  The models had very little overlap
with the data, but rather looped around them, approaching only
the M32 data point.  Thus, the only way to reproduce the data was by
combining the SSP models and then specifically picking out the
composite populations which overlay the data.
Figure~\ref{fig:newsbfcol} updates this comparison using the present
set of models, revising the extinction estimates as in SBF-II, and
including the \vbar\ data presented in \S\ref{sec:obs}.
The locus of the Worthey models, as in Blakeslee \etal\ (1999a),
is shown for comparison. 

The new models provide a more natural explanation of why the mean \vbib\
for the galaxies is $\sim\,$2.4 with a relatively small dispersion.  As the
model metallicity increases, the \vbib\ colour gets redder (though bluer as
the age increases at fixed [Fe/H]), until the metallicity approaches
roughly half solar and higher, where $\vbib$ remains near 2.4.  The models
plotted in Figure~\ref{fig:newsbfcol} use a Salpeter IMF ($\mu=1.3$), but
the form of this plot is not very sensitive to the IMF.  At the
metallicities appropriate to elliptical galaxies, \vbib\ gets bluer by 
about 0.08 mag, and \ibkb\ changes by only about 0.03 mag, when the 
slope of the unimodal IMF is increased to $\mu=2.3$.  As before, a bimodal
IMF with $\mu\approx2$ gives results similar to Salpeter.

This figure is, however, very sensitive to the underlying
isochrones used in the modeling.  As discussed more in the following
sections, the gross differences between the present model predictions and
those of Worthey are mainly due to the different sets of isochrones used.
Liu \etal\ (2000) also find that their models, which use the stellar
tracks of Bertelli \etal\ (1994), are much closer to the data in 
this diagram than the Worthey models are.
For our models, the discrepancy with the Worthey predictions becomes
greater when the Bertelli \etal\ (1994) isochrones are used:
the model points in Figure~\ref{fig:newsbfcol} would be shifted even further
to the left and would exhibit considerably more scatter.

Although our new models help us to understand the observed galaxy
fluctuation colours better,
composite population modeling still appears necessary
for the galaxies in Figure~\ref{fig:newsbfcol} with $\vbib\lta2.3$.
We take up this problem in the following section.

\section{Theoretical Distance Calibrations}
\label{sec:theorydists}

One motivation for this work was to derive the best possible theoretical
calibration of the SBF distance method, particularly in the $I$ and
near-IR bands.  
One complication is that our new SSP models do not generally define simple
linear relations between $\Mbar$ and colour.  This alone does not
indicate disagreement with observations, as actual galaxies are not
expected to be homogeneous simple stellar populations.  However, it
becomes necessary to combine the models in such a way as to reproduce
the observed distribution of integrated and fluctuation colours
(distance-independent quantities) before fitting for a distance
calibration.

\subsection{Recipe for the Composite Populations}

We wish to combine the homogeneous SSP models into composite systems in
such a way as to mimic, at least crudely, the evolution of an
elliptical galaxy.
We first group the SSP models into three bins according to 
metallicity: (1) metal-poor, (2) intermediate, and (3) metal-rich,
with each bin containing two of the six SSP model metallicities.
We then randomly choose one SSP model from each bin,
or `metallicity component,' subject to the following age restrictions:
\begin{eqnarray}
   14.1\hbox{ Gyr} &\le\;\; \hbox{age(1)} \;\;\le& \hbox{17.8 Gyr} \nonumber\\
    8.9\hbox{ Gyr} &\le\;\; \hbox{age(2)} \;\;\le& \hbox{17.8 Gyr}
\label{eq:ages} \\
    4.0\hbox{ Gyr} &\le\;\; \hbox{age(3)} \;\;\le& \hbox{17.8 Gyr} \,. \nonumber
\end{eqnarray}
We use all 14 isochrone ages of 4\,Gyr or more
calculated by Girardi \etal\ (2000).
This includes all the ages listed in Table~\ref{tab:sbf_un1.3}
plus ages of 4.5, 5.6, 7.1, and 8.9~Gyr.  

The three components are then combined with weights \hbox{$f_1,f_2,f_3$},
which are random numbers normalized so that the composite models have in
the mean contributions of 10\% from the low-metallicity bin, 40\% from the
intermediate bin, and 50\% from the high-metallicity bin. We experimented
with other weightings, but kept these for reasons given in the
following sections.  The fluctuation luminosity for the composite
population is calculated as
\begin{equation}
\lbar = {{f_1{\cdot\,}\langle{L_1^2}\rangle + f_2{\cdot\,}\langle{L_2^2}\rangle
+ f_3{\cdot\,}\langle{L_3^2}\rangle}
\over{f_1{\cdot\,}\langle L_1\rangle + f_2{\cdot\,}\langle L_2\rangle +
 f_3{\cdot\,}\langle L_3\rangle}}
\;,
\end{equation}
where by definition
$\langle{L_i^2}\rangle = \langle L_i\rangle{\,\cdot\,}\lbar{_i}$.

We considered incorporating the additional requirement age(1) $\ge$ age(2)
$\ge$ age(3), as would necessarily result from pure self-enrichment models.
Lifting this restriction in favour of Eqs.\,(\ref{eq:ages}) is meant to
allow for the possibility of merging/accretion or late infall of metal-poor
gas.  This extra freedom in the ages increases the scatter in the
\Mbar-\vi\ relations of the following section by $\sim\,$10\% and affects
the slopes at the $\sim\,$10\% level, but does not significantly affect the
zero~points at our fiducial $\vi=1.15$ colour.

Finally, we select out the composite models having the same (within
3$\,\sigma$) fluctuation colours as the galaxy data.  In practice this
includes all composite models with $\ibkb > 3.65$ and $\vbib > 1.95$.
About 42\% of the models meet these selection criteria.  Overall, the
mean ages for the metal-poor, intermediate, and metal-rich components
are 15.9, 13.0, and 9.3~Gyr, respectively; after culling according to
the fluctuation colours, the mean ages become 16.1, 12.9, and 7.6~Gyr,
respectively.  Among the models in this `culled sample,' 12.9\% have
the intermediate metallicity component older than the metal-poor
component, 11.6\% have the metal-rich component older than the
intermediate component, and only 2.8\% have the metal-rich component older
than the metal-poor one.  These `age-overlap' percentages are decreased
from 14\%, 22\%, and 7\%, respectively, in the pre-selection sample.

The fluctuation-colour selection therefore drives the metal-rich component
towards younger ages and significantly decreases the fraction of composite
models with `age inversion' (more metal-poor stars being younger than richer ones).
It may be tempting to conclude, speculatively, that this result favours
self-enrichment models and that the accretion of low-metallicity gas-rich
material (and subsequent formation of young, metal-poor stars) is not an
important factor in the luminosity grown of ellipticals.  This is similar
to the conclusions of Trager \etal\ (2000b).  However, our composite
population scheme is fairly ad~hoc and the model/data comparison
uses only two observables (two fluctuation colours).  While the results are
intriguing, firm conclusions on the evolution of early-type galaxies
await more exhaustive analyses with larger data~sets.

\subsection{Optical SBF Distances}
\label{ssec:optcals}

\begin{figure}
\epsfig{figure=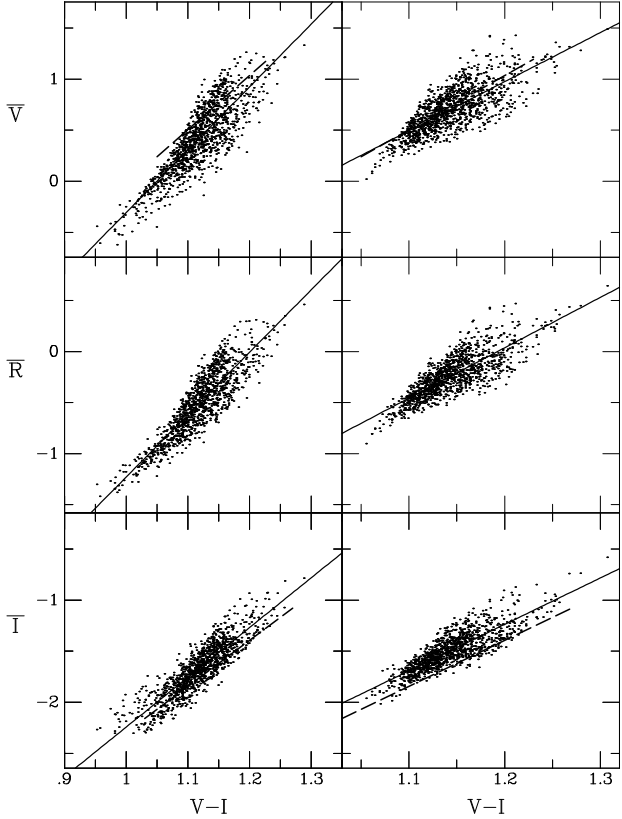, height=0.61\textwidth, width=0.46\textwidth}
\medskip
\caption{Optical $VRI$ SBF magnitudes are plotted against \vi\ for composite 
populations.  In the left panels, no selection criteria have been applied to the
simulated composite populations; in the right panels, only composite models 
having \vbib\ and \ibkb\ fluctuation colours in the observed range 
for elliptical galaxies are shown.  In each case, 1200 points are plotted,
and the solid lines are least-squares fits to guide the eye.
Note the difference in horizontal scale between the left and right panels.
The dashed lines represent the empirical relations in the $V$ (top)
and $I$ (bottom) bands, from this work and the $I$-band SBF survey,
respectively.
}\label{fig:optcals}
\end{figure}

Figure~\ref{fig:optcals} plots the composite model SBF magnitudes in the
$VRI$ bands against integrated \vi\ colour.  The panels on the left
show models prior to any selection criteria, while the ones on right
have been culled according to their fluctuation colours, as described above.
The least-squares fits to the culled models and the rms scatters in
the fits are given by
\begin{eqnarray}
\overline{M}_V &=\;\; +0.74 \,+\, 4.8\,[\vi - 1.15]\,,& \pm\,0.15\,\hbox{mag} \,, 
\label{eq:vbar} \\
\overline{M}_R &=\;\; -0.21 \,+\, 4.9\,[\vi - 1.15]\,,& \pm\,0.13\,\hbox{mag} \,, 
\label{eq:rbar} \\
\overline{M}_I &=\;\; -1.47 \,+\, 4.5\,[\vi - 1.15]\,,& \pm\,0.10\,\hbox{mag} \,.
\label{eq:ibar}  
\end{eqnarray}
The selection by fluctuation colours has decreased the rms scatters by
28\%, 25\%, and 22\% in $V$, $R$, and $I$, respectively.  The
statistical uncertainties in the slopes of these relations are
$\lta\,$0.5\%, but experimenting with other reasonable metallicity
weightings and age restrictions shows that variations of $\pm\,$12\% are
typical and give a more realistic impression of the uncertainty.  The
zero~points, however, are much more secure, typically varying by
$\pm$0.02\,mag or less (for a Salpeter IMF; changing the IMF slope can cause
variations 4 times greater).

One motivation for the adopted composite population scheme is that it
reproduces the slope of 4.5 observed in the $I$~band (SBF-I).  Thus, we
cannot really claim that these models `predict' the observed $I$-band
slope. Rather, the slope was more of an observational constraint in the
modeling, similar to the \vbib\ and \ibkb\ fluctuation colours. 
For instance, decreasing the mean percentage of metal-poor stars from 10\%
to 5\% increases the $I$-band slope by about 8\% and decreases the $V$
slope so that the relation is steeper in $I$, contrary to what is
observed.  Thus, metal-poor stars appear to be a small but important
component of early-type galaxies.  
Part of the reason the model slopes match the observed
ones is that the colour distribution of the composite models,
$1.05\lta\vi\lta1.25$, is similar to that of the SBF survey galaxies.
The $I$~slope is flatter for bluer populations, as discussed in
\S\ref{ssec:sbfcephs} above, and becomes steeper at the red end (see
\S\ref{ssec:hstcal} below).

The $V$-band calibration of Eq.\,(\ref{eq:vbar}) agrees well with the
observed calibration from Eq.\,(\ref{eq:vbempcal}).  The zero~points
differ by just 0.03\,mag.  We note, however, that the adopted empirical
zero point relies on the group tie between SBF and Cepheids, which is
more secure both in a statistical sense and in the sense that the 
calibrators are typical of the sample galaxies.
Had the `direct' tie, based on fairly uncertain SBF measurements in spiral
bulges, been adopted in \S\ref{ssec:vemp}, 
the empirical zero~point ($\Mvz=+0.65$) would 0.09\,mag brighter 
than the zero point of Eq.\,(\ref{eq:vbar}).

The empirical $I$ zero point $\Miz=-1.62$ adopted in \S\ref{ssec:vemp} is
0.15\,mag brighter than the Eq.\,(\ref{eq:ibar}) calibration.  A similar
offset was found in \S\ref{ssec:sbfcephs} when comparing \MbI\ for M32
to the SSP models.  Had we adopted the `direct' Cephe\-id tie from SBF-II,
the discrepancy would be 0.27\,mag.  We take this as additional evidence in
favour of the group calibration.  Further, we note that the dynamical distance
to the NGC\,4258 water maser (Herrnstein \etal\ 1999) indicates that the
Cepheid distance scale may be too long by 0.2\,mag (Maoz \etal\ 1999).  If
so, we would achieve excellent agreement between the group $I$-band SBF
calibration and our theoretical calibration.  The theoretical $V$-band
calibration would then be $\sim0.2\,$mag too bright, i.e., the \vbib\ colours of
our composite models are too blue by about 0.15\,mag.  It is possible to
`fix' this situation by weighting them heavily towards the high metallicity
component, but then the \vi\ colours become redder than the SBF survey
galaxies by $\sim\,$0.08\,mag and the $I$ slope increases to $\sim6$.

Finally, we note that our theoretical \MbI\ calibration is
$\sim\,$0.35\,mag fainter than the calibration from the Worthey (1994)
models (see SBF-I; Blakeslee \etal\ 1999a).  At solar metallicity, the \vi\
colours of the Worthey models are $\sim\,$0.07\,mag redder than ours.
These differences are mainly due to the different isochrone sets.  As
discussed in detail by Charlot \etal\ (1996), the mean luminosity of the
RGB stars is about 25\% greater in the isochrones used by Worthey.  For the
integrated light, this is mostly offset by the greater quantity of stars on
the Padua RGB.  However, even if these two competing effects completely
cancel for the integrated light, the luminosity variance from the Worthey
RGB will be 0.25\,mag brighter.  G.~Worthey's stellar population 
page\footnote{http://astro.sau.edu/$\sim$worthey} on
the World Wide Web provides the option of replacing his original isochrones
with the Bertelli \etal\ (1994) isochrones. Using this option makes the
Worthey models at solar metallicity bluer in \vi\ by nearly 0.1\,mag and
fainter in \MbI\ by $\sim\,$0.50\,mag; the resulting calibration is about
0.2\,mag fainter than our \MbI\ calibration (which uses the more recent
Padua isochrones and a completely different set of transformations).
Yet, the new Liu \etal\ (2000) models use the Bertelli \etal\ tracks
and obtain a zero point similar to the original Worthey (1994) one.
Table~\ref{tab:izeros} provides a summary scorecard of the various
empirical and theoretical $I$~SBF calibrations.

This discussion illustrates how the combination of integrated colours
and SBF magnitudes can provide important new constraints for stellar
evolution.  At the same time, the sensitivity of SBF to the finer
details of RGB evolution means that any theoretical calibration 
will at this point be inherently uncertain.

\begin{table}
\centering
\begin{minipage}{82mm}
  \caption{Comparison of $I$-band SBF Zero Points}
\label{tab:izeros}
\def\midofcol#1{\hfill\hskip -20pt plus 1000pt#1\hskip -12pt plus 1000pt\hfill}
\newdimen\digitwidth
\setbox0=\hbox{\rm0}
\digitwidth=\wd0
\catcode`?=\active
\def?{\kern\digitwidth}
  \begin{tabular}{ccl}
\hline
?$\overline M^0_I\,$\rlap{\footnote[1]{$I$-band SBF zero point evaluated at
$\vi=1.15$ using a fixed slope of 4.5. Formal uncertainties
on the empirical numbers are $\sim\pm0.1$\,mag (see SBF-II and Ferrarese \etal).}} & 
Method\rlap{\footnote[2]{`Spiral' for the empirical direct SBF-Cepheid tie via spiral bulges;
`group' for the empirical SBF-Cepheid tie via galaxy groups;
`theory' for stellar population models.}}
& \midofcol{Source} \\
\hline                                                
$-$1.74 & spiral & SBF-II  \\
$-$1.79 & spiral & Ferrarese \etal\ 
(2000)\rlap{\footnote[3]{Ferrarese \etal\ used the same SBF and Cepheid
distances as SBF-II, but a different weighting scheme.}} \\
$-$1.54 & spiral & SBF-II, adjusted for NGC\,4258\rlap{\footnote[4]{Resulting 
zero point if the Cepheid scale is revised according to the Herrnstein \etal\ (1999)
NGC\,4258 maser distance.}} \\
$-$1.62 & group  & SBF-II \\
$-$1.42 & group  & SBF-II, adjusted for NGC\,4258\rlap{$^d$} 
\vspace{3pt}\\
$-$1.47 & theory & This work, Eq.~\ref{eq:ibar} \\
$-$1.81 & theory & SBF-I, Worthey (1994) models \\
$-$1.25 & theory & Worthey Web page, Padua
option\rlap{\footnote[5]{`Padua isochrones' used here are
those of Bertelli et\,al.~(1994).}}\\
$-$1.79 & theory & Liu \etal\ (2000)\\
\hline
\end{tabular}
\vspace{-10pt}
\end{minipage}
\end{table}

\begin{figure}
\epsfig{figure=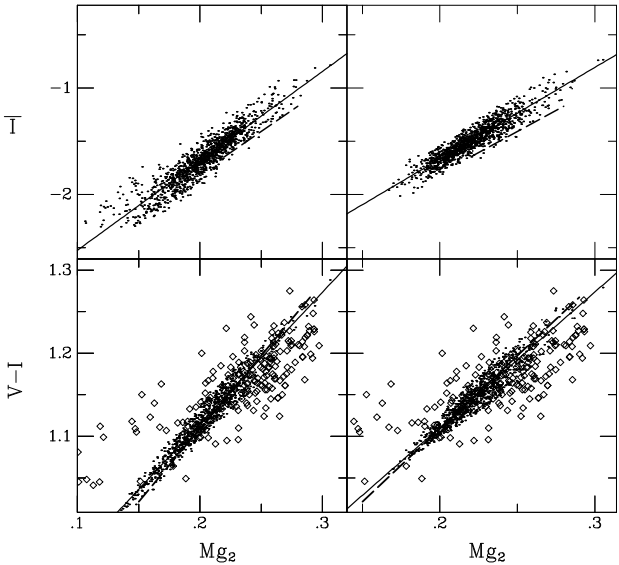, height=0.42\textwidth, width=0.46\textwidth}
\medskip
\caption{Model $I$-band SBF magnitude \MbI\ and \vi\ colours
are plotted against \mgii\ for our composite models.  As in
Figure~\ref{fig:optcals}, the left panels show models without any
selection criteria, and the right panels show models selected based on
fluctuation colours, with the solid lines being least-squares fits.
Note the difference in horizontal scale between the left and right panels.
The dashed line in the top panels represents
the empirical relation from Thomsen \etal\ (1997).  The dashed line in
the lower panels shows the relation between \vi\ and \mgii\ obtained by
tying together the two separate \MbI\ calibrations; for visibility,
this line is extended well beyond its true validity limits.
Data for nearly 200 galaxies (open diamonds) in common between the SBF
and SMAC surveys are shown for comparison; their \mgii\ values have
been shifted by $-0.04$ as an overall estimate of the correction
for aperture differences.
}\label{fig:mg2cal}
\end{figure}

\subsection{Mg$_2$ versus \vi\ for the Calibration}

The $I$-band SBF distance indicator has also been calibrated against the
Mg$_2$ absorption index (Thomsen \etal\ 1997).
This calibration,
using \mi\ data from Sodemann \& Thomsen (1995) and the Mg$_2$ 
measurements of Davies, Sadler, \& Peletier (1993),
is based solely on the internal gradients in NGC\,3379.
It is only strictly applicable over the range
$0.24<\hbox{Mg}_2<0.28$.  The top panels of Figure~\ref{fig:mg2cal} show
the corresponding relation for our composite models.  The relation in
the top right panel is
\begin{equation}
\overline{M}_I \;=\; -1.50 \,+\, 8.6\,[{\rm Mg}_2 - 0.22]\,,\;\; \pm\,0.07\,\hbox{mag} \,.
\label{eq:ibarmg2}
\end{equation}
The slope agrees well with the Thomsen \etal\ calibration, but the
zero~point is again 0.15\,mag too faint.  Although the scatter is small,
calibrating the observations is much more difficult.  It requires a
well-orchestrated combination of imaging and spectroscopy, and absorption
measurements are difficult well outside the bright galaxy centre.
Also, there is the unresolved (and unmodeled) issue over whether the colours,
and SBF magnitudes for that matter, should follow the Fe
abundance or the Mg overabundance.

The consistency between the Mg$_2$ and \vi\ calibrations is illustrated in
the lower panels of Figure~\ref{fig:mg2cal}. We can tie these two
quantities together via their respective \MbI\ calibrations, which for \vi\
is based on $\sim\,$150 galaxies and for \mgii\ is based on one galaxy. The
resulting `empirical' line, extended well beyond the true limits of
applicability, almost perfectly overlays the model predictions.  However,
the agreement may be somewhat coincidental.  A shift of only 0.02\,mag in
the Mg$_2$ data for NGC\,3379 (as, for instance, in the measurements by
Vazdekis \etal\ 1997) would eliminate the offset in the top panel of
Fig.~\ref{fig:mg2cal} and cause an offset in the lower panel.  For a
further comparison we show in the bottom panel galaxies in common between
the SBF and SMAC (Smith \etal\ 2000; Hudson \etal\ 2000) surveys.  The SMAC
Mg$_2$ values have been shifted by $-0.04$ mag as an approximate allowance
for aperture effects.  The galaxy points follow a shallower slope than the
models, perhaps because of the increasing Mg enhancement for redder, more
luminous ellipticals.

\begin{figure}
\epsfig{figure=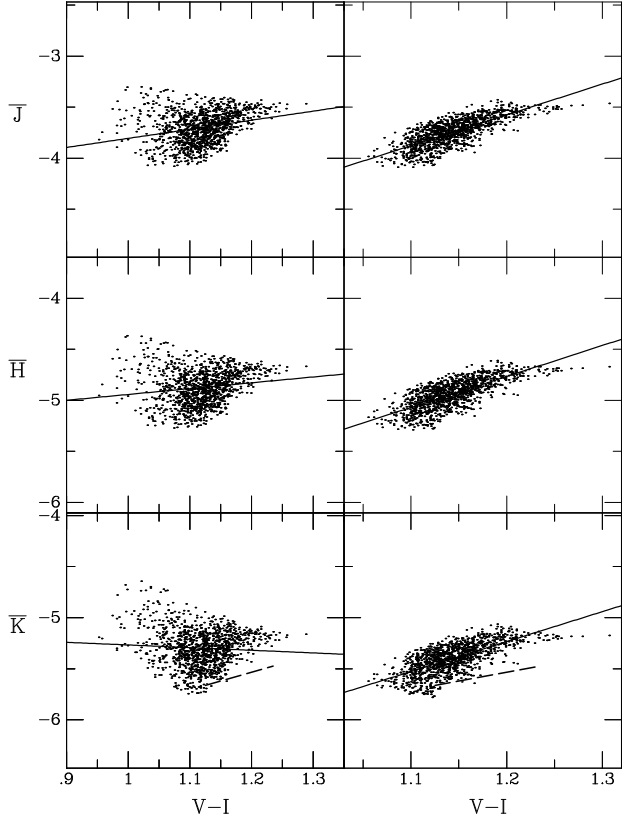, height=0.61\textwidth, width=0.46\textwidth}
\caption{Similar to Figure~\ref{fig:optcals} except for the near-IR bandpasses.
Again, note the horizontal scale difference between the left and right panels.
The dashed line in the bottom panels is the empirical two-parameter
group calibration from table\,7 of Jensen \etal~(1998).}\label{fig:ircals}
\end{figure}

\subsection{Near-Infrared SBF Distances}

Figure~\ref{fig:ircals} shows the composite model relations between \vi\ colour
and the near-IR SBF magnitudes.  The calibrations derived from the simple
least-squares fits are
\begin{eqnarray}
\overline{M}_J &=\;\; -3.72 \,+\, 3.0\,[\vi - 1.15]\,,& \pm\,0.09\,\hbox{mag} \,, 
\label{eq:jbar} \\
\overline{M}_H &=\;\; -4.92 \,+\, 3.0\,[\vi - 1.15]\,,& \pm\,0.10\,\hbox{mag} \,, 
\label{eq:hbar} \\
\overline{M}_K &=\;\; -5.38 \,+\, 2.9\,[\vi - 1.15]\,,& \pm\,0.10\,\hbox{mag} \,.
\label{eq:kbar}  
\end{eqnarray}
Again, the slopes show variations at about the 12\% level for different
(reasonable) composite modeling assumptions, while the zero points remain
fixed at the $\pm\,$0.02\,mag level.
The selection by fluctuation colours decreases the rms scatters
about Eqs.\,(\ref{eq:jbar})-(\ref{eq:kbar}) by $\sim\,$70\% in each case.
Thus, a measurement of the actual scatter in a large and diverse sample
of early-type galaxies could provide insight into the variety
of the stellar mixtures in these galaxies.

Our model $K$-band zero point at $\vi=1.15$ is 0.24--0.36 mag fainter
than the empirical one (depending on whether the group or direct Cepheid
tie is used), which is based on measurements by Jensen \etal\ (1998).
We should note that the $K$-band data are more weighted towards the red
than the $I$-band survey data, so a better fiducial colour might be
$\vi=1.2$, in which case the zero~point discrepancy is 0.3--0.42\,mag.
Interestingly, the former Padua isochrones (Bertelli \etal\ 1994),
because of their more infra\-red luminous AGB,
make the model $K$ zero~point brighter by 0.5\,mag or more, i.e., about
0.2\,mag brighter than the empirical zero point.  However, the agreement
becomes much worse in the $I$~band, with the theoretical zero point
being 0.4--0.5 mag too faint because of the cooler RGB.
Again, all the same caveats apply as in \S\ref{ssec:optcals}
with regard to the sensitivity of these calibrations to stellar evolution.
Clearly we will have learned much by the time the models reproduce
the SBF data in all bands.

It is noteworthy that the slopes of Eqs.\,(\ref{eq:jbar})--(\ref{eq:kbar})
are positive, i.e., in the same sense as the slopes for optical SBF
(although the $K$-band slope is negative before the selection by
fluctuation colours).  This is in the sense opposite that given by 
either the Worthey models or the Liu \etal\ (2000) models, which both
predict a brightening of the near-IR SBF magnitudes for redder populations.
The dimming of our model near-IR SBF magnitudes in redder populations
agrees better with observations (Jensen \etal\ 2000, in preparation). 
In particular, we disagree with the conclusion by Liu \etal\ (2000) 
that the $K$-band data imply that ellipticals all have roughly solar
metallicity but a large spread in age.  In the context of our models,
the observed trend could well be due to metallicity variations.

Redder than $\vi=1.2$, Figure~\ref{fig:ircals} shows a flattening of
the predicted relations, so that a constant \Mbar\ calibration in the
near-IR would work best for
the reddest ellipticals.  Interestingly, just the opposite happens in
the $I$-band, as we discuss in the following section.

\subsection{SBF with WFPC2}
\label{ssec:hstcal}

Ajhar \etal\ (1997) measured SBF magnitudes in the
WFPC2 F814W bandpass for 16 early-type galaxies
observed with {\it HST}.
Although F814W is similar to Cousins $I$, the data gave
a steeper slope of $6.5\pm0.8$ for the calibration against \vi.
In an effort to understand this, we have computed SBF magnitudes
for our models in the WFPC2 filter system using the empirical
and synthetic photometric transformations from Holtzman \etal\ (1995).
Unfortunately, the empirical transformations are only determined for
stars with $\vi\lta1.4$.  For redder stars, including higher metallicity
RGB stars (the most important ones for us), they derived synthetic
transformations using template stellar spectra and model bandpasses.  It
is this sort of theoretical transformation that we have sought to avoid
in using the new Vazdekis models, but here we are forced to it.

\begin{figure}
\epsfig{figure=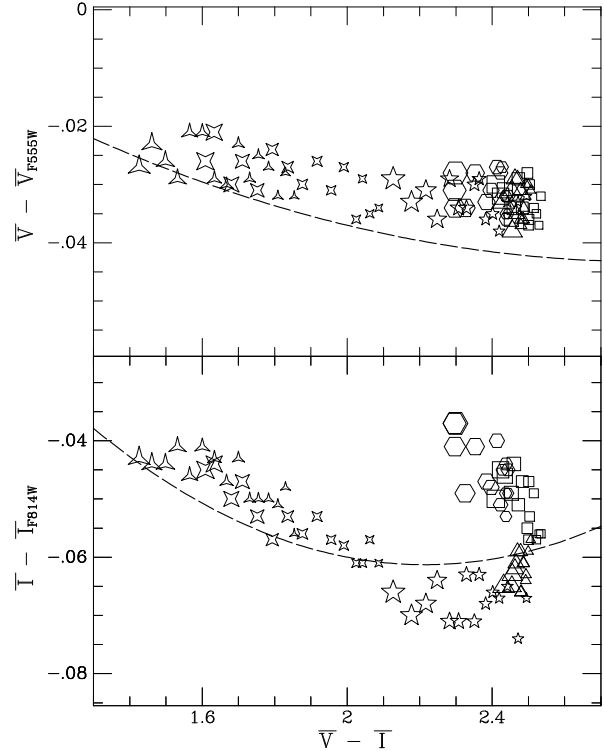, height=0.56\textwidth, width=0.44\textwidth}
\medskip
\caption{A comparison of the synthetic stellar transformations from WFPC2
to Johnson-Cousins bandpasses (Holtzman \etal\ 1995 -- dashed lines)
with the SBF transformations `measured' from the SSP models (symbols as in
Figure~\ref{fig:newcals}).  Small but systematic errors will be made in
assuming that the stellar transformations will apply to the SBF
measurements.  The true errors will be larger than implied by this figure,
as the Holtzman \etal\ transformations were built into our code for
computing the WFPC2 magnitudes of the constituent model stars, but actual
stars will deviate from these transformations.  Because we were forced to
rely heavily on synthetic transformations (shifted in a fairly arbitrary
way into agreement with the empirical transformations), our model SBF
magnitudes for the WFPC2 filters are more uncertain than for the standard
bandpasses.  }\label{fig:hstdiffs}
\end{figure}

\begin{figure}
\epsfig{figure=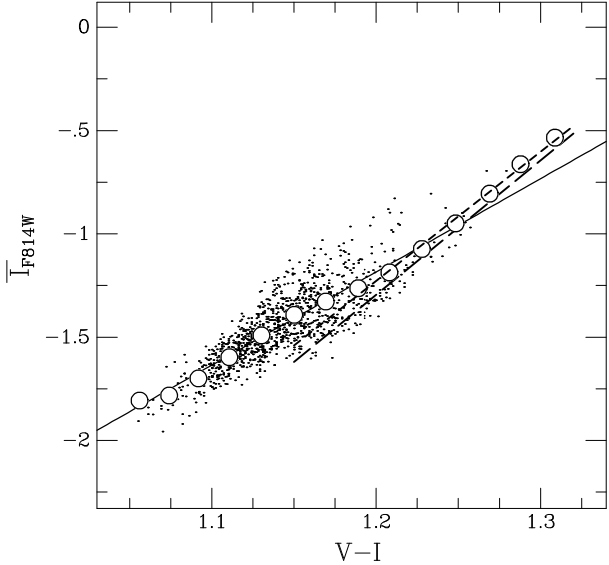, height=0.42\textwidth, width=0.45\textwidth}
\medskip
\caption{The F814W SBF magnitudes are plotted against \vi\ for composite
models (after the fluctuation-colour selection).  Again, 1200 individual
composite models are plotted with a linear fit (solid line), but the open
circles represent averages in \vi\ steps of 0.02\,mag for over 80,000 model
realizations.  A linear fit to the open circles with $\vi>1.18$
(short-dashed line) has a slope in close agreement with the slope of the
F814W SBF calibration derived by Ajhar \etal\ (1997) (long-dashed line).
The vertical offset between the models and the empirical line is 0.06\,mag,
or 0.09\,mag less than the corresponding offset for Cousins $I$, but this
small discrepancy may be due at least in part to systematic errors in the
synthetic stellar transformations.  }\label{fig:F814cal}
\end{figure}

As Holtzman \etal\ discuss, their empirical and synthetic
transformations do not always agree.  However, we noticed that the two
types of $I$--F814W transformation intersect at $\vi=1.31$.  For
consistency, we then shifted the $V$--F555W and $R$--F675W synthetic
transformations by $\sim\,$0.023\,mag into agreement with the empirical
ones at $\vi=1.31$.  We used the empirical transformation for
model stars bluer than this and the synthetic one for redder model
stars.  After transforming all the stars, we calculate the SBF
magnitudes.  Figure~\ref{fig:hstdiffs} shows the differences between the
standard and WFPC2 $V$ and $I$ SBF magnitudes as a function of \vbib,
and compares these differences to the stellar transformations.  Even
though these are precisely the transformations used for the model stars
in this colour range, there are systematic (though small) differences
with the final SBF transformations.  This should be no surprise:
the spectrum of the fluctuations is quite different from a stellar spectrum,
but it is a point often casually overlooked.

Figure~\ref{fig:F814cal} shows the resulting relation between $\overline
M_{{\rm 814W}}$ and \vi\ for the culled sample of composite models.  It
has the same overall 4.5 slope as in the standard $I$ band, and is
shifted fainter by 0.06\,mag (but this result depends on the uncertain
synthetic transforms).  However, there is a steepening of the
relation starting around $\vi=1.2$.  To illustrate, we compute the
running average of $\overline M_{{\rm 814W}}$ in \vi\ steps of 0.02 and plot
the results in the figure.  The Ajhar \etal\ (1997) calibrating galaxies
are distributed from $\vi=1.15$ to $\vi=1.31$, with 75\% being redder
than 1.20.  This is because the WFPC2 images show only the red galaxy centres.
The slope of the running averages at $\vi>1.18$ is 6.2 [and
changes by $\pm0.3$ as the starting \vi\ is changed by $\pm0.02$],
in good agreement with the Ajhar \etal\ result.  Thus, these new models
for the first time resolve the puzzle of the steeper F814W slope: it
boils down to a slightly nonlinear relation coupled with differences in
sample selection.~~~~

\section{The Fluctuation Star Count}\label{sec:nbar}

Recently, SBF-IV introduced the parameter \nbar\ as the ratio, expressed 
in magnitudes, of the total apparent flux from a galaxy to the flux in the
fluctuation signal.  This ratio is independent of distance, photometric
calibration, and extinction; it corresponds to the total luminosity 
of a galaxy in units of a typical giant star within that galaxy:
\begin{equation}
\nbar \,=\, \mbar - m_{\rm tot} \,=\,
    +2.5\,\log\left[L_{\rm tot} \over \lbar\right] \,,
\label{eq:nbardef}
\end{equation}
where $m_{\rm tot}$ is the total apparent magnitude and $L_{\rm tot}$
the total luminosity.  Put another way, if all stars in a galaxy were of
magnitude \mbar, then \nbar\ would be 2.5 times the logarithm of the total
number of stars. SBF-IV therefore called \nbar\ the ``fluctuation star
count,'' but it is known to the authors for convenience as the ``Tonry
number.''

SBF-IV showed that \nbar\ can be used as an alternative calibration
in measuring SBF distances.  The calibration they derived,
adjusted to our group-based zero point (\S\ref{ssec:vemp}), is
\begin{equation}
  \Mi = -1.62 + 0.14(\Nbar-20) \,.
  \label{eq:nbarmbar}
\end{equation}
Although this introduces covariance between \mi\ and \Mi,
the final distance modulus is actually 14\% less sensitive to
errors in \mi.  Moreover, the necessity of measuring \vi\
to an accuracy of 0.022 mag in order to limit the distance
error from photometry to $\lta0.1\,$mag is replaced by the much 
less stringent requirement of measuring $m_{\rm tot}$ to an accuracy
of 0.7\,mag.  Similarly, the sensitivity to errors in the Galactic
extinction is cut nearly in half for the \nbar\ calibration.

However, SBF-IV cautioned against dropping \vi\ altogether in favour
of \nbar.
The \vi\ calibration is based purely on stellar population properties,
whereas the \nbar\ calibration essentially relies on the fundamental
plane (FP) of elliptical galaxies and may be subject to systematic effects,
such as environmental influences or the presence of a disk component.
SBF-IV showed that \nbar\ has a tight correlation with stellar 
velocity dispersion for SBF survey ellipticals:
\begin{equation}
  \log(\sigma) = 2.22  + 0.10\, (\Nbar - 20)\,,
  \label{eq:nbarsig}
\end{equation}
where $\sigma$ is in \kms, and the scatter is 0.070 dex.
The connection with the FP becomes transparent
if we invert this relation to get
\begin{equation}
 L_{\rm tot} \;=\;  10^8 \lbar \,
\left(\sigma \over 165{\,\rm km\, s^{-1}}\right)^{4.0} .
\label{eq:fabernbar}
\end{equation}
Eq.\,(\ref{eq:fabernbar}) superficially resembles the classical
Faber-Jackson relation (Faber \& Jackson 1976), but it contains an
implicit correction for the mass-to-light ratio via the \lbar\ term,
with its strong  stellar-population dependence. This explains why the
scatter is similar to that of the inverse FP (e.g., Hudson \etal\ 1997),
and less than half that of Faber-Jackson.

In order to model the behaviour of \nbar, we must assume some relation
between the galaxy luminosity, or mass, and the stellar population.
Perhaps the most common form this takes is the Mg$_2$-$\sigma$ relation
(Dressler \etal\ 1987),
conventionally interpreted as a pure mass--metallicity relation 
(e.g., Bender, Burstein, \& Faber 1993; Guzm\'an, Lucey, \& Bower 1993;
but see also Trager \etal\ 2000b).  However, we wish to avoid the use of
Mg$_2$ here because it explicitly brings in the alpha-enhancement issue.

Instead, we simply derive the following relation between the \vi\ colours
of the SBF survey galaxies (SBF-IV) and the velocity dispersions of
these galaxies from the homogenized SMAC sample (Hudson \etal\ 2000):
\begin{equation}
  \vi \,=\, 0.71 + 0.20\log(\sigma) \,.
  \label{eq:visig}
\end{equation}
Details of the comparison of the SBF and FP parameters will be given
by Blakeslee \etal\ (2000, in preparation).  
Next we assume that the dynamical mass of a galaxy is given
by $M=5\sigma^2r_e/G$ (e.g., Bender \etal\ 1992; J{\o}rgensen 1999)
and make the zeroth-order assumption that all ellipticals have
effective radii $r_e=6$ kpc.  Finally, we assume that the initial
stellar mass is half the dynamical mass and then use the model
mass-to-light ratios to derive $L_{\rm tot}$ and \nbar.

\begin{figure}
\epsfig{figure=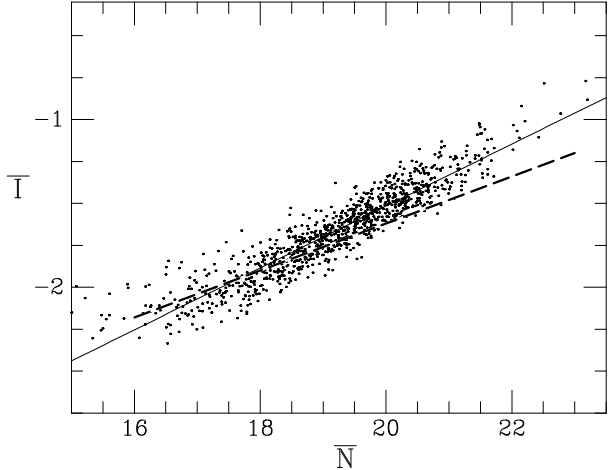, height=0.35\textwidth, width=0.45\textwidth}
\medskip
\caption{\MbI\ for the composite models is plotted against the
Tonry fluctuation number \nbar, which has been computed
assuming an empirical colour--velocity~dispersion relation.
The solid line is a least-squares fit to the models, and the thick
dashed line is the empirical relation from SBF-IV (Tonry \etal\ 2000b).
\label{fig:nbar}}
\end{figure}

Figure~\ref{fig:nbar} illustrates that our crude set of
assumptions produces an \Mi-\nbar\ relation very similar to the
observed one.  The slope of the composite model relation is 0.18,
while the empirical slope is $0.14\pm0.04$, and the scatter is
0.095 mag (the slope and scatter would change to 0.20 and 0.074 mag,
respectively, if we culled as in the previous section).
The zero points of the model and empirical relations are also
similar in Figure~\ref{fig:nbar}, but this of course depends on
the values assumed for $r_e$ and the fraction of the mass in
stellar material.  We have provided only a fairly schematic
explanation, based on scaling laws, of why the \nbar\ distance
calibration works.  A true theoretical calibration of this method
must come from further developments in the semi-analytic modeling
of early-type galaxy formation 
(e.g., Cole \etal\ 1994; Baugh, Cole, \& Frenk 1996; 
Kauffmann \& Charlot 1998).

\section{Summary and Conclusions}\label{sec:conclusions}

We have presented new $V$-band SBF measurements for five early-type
galaxies in the Fornax cluster.  Along with Virgo and Leo $\vbar$ data
from Tonry \etal\ (1990), these have been used to derive an empirical,
Cepheid-calibrated $\MbV$-\vi\ distance indicator.  The slope of this
relation is $5.3\pm0.8$, similar to the $\MbI$-\vi\ empirical slope of
4.5 (SBF-I).  We find a mean fluctuation colour $\vbib=2.4$ for our
fairly homogeneous set of ellipticals, with a dispersion of only
0.15\,mag, comparable to the measurement error.

New stellar population models, updated from those of Vazdekis \etal\
(1996) with the isochrones of Girardi \etal\ (2000) and a new set of
empirical transformations, have been employed to calculate SBF
magnitudes and integrated colours for six different metallicities and a
wide range of ages.  The age-metallicity degeneracy in the optical
\Mbar-colour relations is less for these models than for the models of
Worthey (1994), although there is still a high degree of degeneracy in
the relation against \mgii\ at intermediate and high metallicities.  In
the near-IR bands, the model SBF magnitudes have much shallower
dependences on colour/metallicity.  Below solar metallicity, they depend
solely on age, while the integrated colour depends almost solely on
metallicity.  Thus, the \MbH\ vs.\ $(I{-}K)$ plot completely breaks the
age-metallicity degeneracy for the lower metallicity models.  This could
be an important new avenue in stellar population studies.

We have examined the sensitivity of our SBF predictions to changes in
the isochrones, theoretical-to-observational transformations, and the
IMF.  We find that using the former set of Padua isochrones makes the
SBF magnitudes 0.1--0.2\,mag fainter in the optical and about 0.5\,mag
brighter in the near-IR.  Using theoretical transformations instead of
our empirical ones can change the \MbI\ predictions by nearly 0.5\,mag
and \MbH\ by up to 0.7\,mag; the changes for the other bands are at the
0.1--0.2 mag level. Thus, the isochrones and the transformations have
about equal importance for the SBF results.
On the other hand, different IMFs yield fairly similar \Mbar-colour
relations in the optical because the IMF effects are largely degenerate
with age and metallicity, as noted by Worthey (1993a).  This is less
true in the near-IR, where the steeper IMF still makes the model colour
redder (like metallicity) and \Mbar\ fainter (like age), but age and
metallicity are no longer degenerate.

The \MbV\ and \MbI\ magnitudes of our low metallicity models have been
tested against observations for 19 globular clusters.  The models
agree fairly well, especially if younger ages are taken at higher
metallicities, with the data in both zero point and metallicity
dependence, which is very weak in the $I$-band.  Model calculations at
lower metallicities and SBF measurements of higher metallicity globulars
(if possible) would allow for much better tests.  At any single metallicity,
the models predict an anticorrelation between \vbib\ fluctuation colour
and integrated \vi\ colour: increasing the age makes \vbib\ bluer and \vi\ redder,
although they both become redder with increasing metallicity. 
The models further imply that the observed spread in \vbib\ at a fixed \vi\
should be interpreted as a spread in age among the globulars.
However, some of the globulars are too blue in \vbib\ 
(and thus too old) to be match models.
This could in part be due to errors in the extinction estimates.  
At high metallicities, the models predict $\vbib\approx2.4$, 
in good agreement with the observations of elliptical galaxies.

Because the single-metallicity, single-age models do not yield
completely monotonic relations between \Mbar\ and colour, we adopt a
Monte Carlo approach to making composite population models from three
metallicity bins with different (fairly loose) restrictions on age.  We
then select out the models that reproduce the observed
(distance-independent) fluctuation colours of early-type galaxies.  This
procedure reproduces well the \vi\ colour range of elliptical galaxies
and the empirically well-determined slope of \MbI\ against \vi.  If we
reduce the mean number of metal poor stars in the models from 10\% to
5\%, the $I$-band SBF slope becomes marginally too steep, and steeper
than the $V$-band slope, contrary to observations.  This is due to the
flattening of the \MbI-\vi\ relation in bluer populations, seen in
both the data and our models, and is a palpable indication of the
presence of metal-poor stars in early-type galaxies.  
Similarly, we find that the $I$ slope becomes steeper for redder
composite populations, those with few metal-poor stars.  This explains
the steep slope found by Ajhar \etal\ (1997) for {\it HST} F814W observations
of elliptical galaxy centres, which are usually quite red.

Although the slopes of the model \Mbar-\vi\ relations are fairly
sensitive to the details of the composite model prescription, the zero
points at $\vi=1.15$ are not.  The zero point of the model \MbV\
calibration is only 0.03 mag brighter than the empirically
determined zero point from \S\ref{sec:obs}.  If the `direct' (but
observationally more precarious) tie of SBF to the Cepheid distance
scale via spiral bulges were adopted instead of the group tie, then the
theoretical calibration would be 0.09 mag fainter than the empirical
one.  The theoretical \MbI\ zero point is 0.15 mag fainter than the
empirical one (or 0.27 mag for the direct tie).  On the other hand, if
the Cepheid distance scale were revised to match the dynamically
measured distance to the water maser in NGC\,4258, then the theoretical
and empirical \MbI\ zero points would agree closely (at the expense of
the empirically less secure \MbV\ calibration).
Our theoretical \MbI\ calibration is $\sim\,$0.35\,mag fainter 
than the calibration given by the Worthey models; this difference
appears to stem from the different isochrones used
in the two sets of models.

The variation of \MbI\ with Mg$_2$ for the models is in good accord with
that obtained by Thomsen \etal\ (1997) from the internal gradients of
NGC\,3379.  Again, the theoretical zero point is 0.15\,mag fainter than
observed.  This relation is predicted to have less scatter than the
calibration against \vi, but the distances would be much more
challenging from an observational point of view, unless the SBF
measurements are restricted to the bright galaxy centres, as for 
example with {\it HST} imaging.  Also, the models do not include
the effects of Mg overabundance on either the linestrength predictions
or the isochrone temperatures, and it is unclear how this would 
affect the theoretical calibration.

Our composite modeling predicts that the slope of the near-IR \Mbar-\vi\
relation will have the same sense as in the optical, in contrast to
the predictions of the Worthey (1994) models and the 
Liu \etal\ (2000) models.  Our predicted relation
then becomes nearly flat for $\vi>1.2$, (just the opposite of the behaviour
seen in the $I$~band).  The slope of our derived relations depends on
our procedure for selecting out composite models based on their
fluctuation colours.  The selection criteria also greatly reduce the
scatter in the near-IR \Mbar-\vi\ relations.  Observational studies of
the true slope and scatter, and further investigation into the range of
fluctuation colours among elliptical galaxies, would be important from
the standpoint of both stellar population and distance studies.  The
$\MbK$ zero point given by our models is 0.24 mag fainter than the
empirically calibrated one (or 0.36 mag fainter with the direct tie).
This discrepancy depends strongly on the temperature structure
of the isochrones, including the AGB.

Finally, we have used mass-colour scaling relations to investigate
the recently-introduced Tonry number \nbar, which describes the
luminosity of a galaxy in units of a typical giant star.
The simple scaling relations combined with the stellar population models
can reproduce the observed dependence of \Mi\ on \Nbar\ from SBF-IV.
However, a true theoretical calibration of this form of the SBF 
method requires more physically-motivated evolutionary models.

Overall, our model SBF predictions are consistent with observations at the
$\sim\,$0.2\,mag level in the optical and at the $\sim\,$0.3\,mag in the
near-IR.  The present uncertainties in the empirical calibrations are
thought to be about a factor of two smaller than this.  These results
therefore present challenges for stellar evolution and population synthesis.
We have noted that SBF magnitudes are acutely sensitive to the distribution
and properties of evolved stars, particularly on the RGB, which in turn are
sensitive to fairly ill-constrained things such as the colour-temperature
scale for the coolest stars, the helium fraction, the prescription for
convection, variations in elemental abundance ratios, diffusion, etc.
However, it is precisely this sensitivity that gives SBF its own
unique vista, as compared to more traditional observables,
and should make multi-band SBF observations an important constraint 
for future stellar population \hbox{modeling}.~ 

\section*{Acknowledgments}
We are grateful to Scott Trager, Guy Worthey, Michael Liu, John Lucey, 
and John Tonry for helpful comments on earlier versions of the manuscript,
and to Harald Kuntschner and Tom Shanks for fruitful discussions.
We thank Guy Worthey for making his stellar population models
accessible via the World Wide Web and the Padua group for
making their isochrones electronically available.
We thank our SBF Survey collaborators John Tonry and Alan
Dressler, as well John Lucey and the SMAC team, for allowing us 
to make use of survey data in the process of publication.
JPB and EAA thank John Tonry for inspiring their interests in SBF.
This work made use of Starlink computer facilities and was supported
at the University of Durham by a PPARC rolling grant in 
Extragalactic Astronomy and Cosmology.

\end{document}

\begin{table*}
 \centering
 \begin{minipage}{170mm}
  \caption{Fornax Galaxy Data}
\label{tab:obs}
\newdimen\digitwidth
\setbox0=\hbox{\rm0}
\digitwidth=\wd0
\catcode`?=\active
\def?{\kern\digitwidth}
  \begin{tabular}{cccccccccccc}
\hline
Galaxy & $v_h$ & $A_V$ & sec$\,z$ & Exp. & PSF & $m_1^*$ & $V{-}I$ & \vbar & $\pm$ & $\vbar{-}\ibar$ & $\pm$ \\
   & (km/s) & (mag) &   & (sec) & (\arcsec) &(mag) &(mag) & (mag) & (mag) & (mag) & (mag)\\
\hline                                                
N1316 &   1760 & 0.07 & 1.01 & 3000 & 1.04 & 35.22 & 1.13 & 32.25 & 0.16 & 2.42 & 0.22 \\
N1344 &   1169 & 0.06 & 1.02 & 2400 & 1.04 & 34.98 & 1.14 & 31.95 & 0.11 & 2.28 & 0.31 \\
N1380 &   1877 & 0.06 & 1.00 & 2400 & 1.03 & 34.99 & 1.20 & 32.33 & 0.12 & 2.54 & 0.19 \\
N1399 &   1425 & 0.04 & 1.10 & 2400 & 0.94 & 34.98 & 1.23 & 32.47 & 0.12 & 2.36 & 0.18 \\
N1404 &   1947 & 0.04 & 1.02 & 2400 & 1.04 & 35.00 & 1.22 & 32.48 & 0.12 & 2.28 & 0.20 \\
\hline
\end{tabular}
\end{minipage}
\end{table*}

\newpage

\begin{table*}
\centering \begin{minipage}{150mm}
\caption{SBF Predictions from New Models for Salpeter $\mu{=}1.3$ IMF}\label{tab:sbf_un1.3}
\newdimen\digitwidth
\setbox0=\hbox{\rm0}
\digitwidth=\wd0
\catcode`?=\active
\def?{\kern\digitwidth}
\begin{tabular}{crccccrrrrrrrr}
\hline
[Fe/H] & Gyr & $B{-}V$ & $V{-}I$ & $J{-}K$ & Mg$_2$ & $\overline M_U$ & $\overline M_B$ & $\overline M_V$ & 
$\overline M_R$ & $\overline M_I$ & $\overline M_J$ & $\overline M_H$ & $\overline M_K$\\
\hline
$-$1.7 &  4.0 &  0.58 &  0.82 &  0.62 &  
0.046\rlap{\footnote[1]{These numbers are uncertain because they require
extrapolation of the Worthey \etal\ (1994) fitting functions.}} &  
	0.86 &$-$0.26 &$-$2.05 &$-$3.03 &$-$3.88 &$-$5.01 &$-$5.95 &$-$6.16 \\
$-$1.7 &  5.0 &  0.59 &  0.83 &  0.61 &  0.048\rlap{$^a$} &  0.95 &$-$0.12 &$-$1.86 &$-$2.81 &$-$3.64 &$-$4.76 &$-$5.69 &$-$5.90 \\
$-$1.7 &  6.3 &  0.61 &  0.85 &  0.61 &  0.050\rlap{$^a$} &  1.03 &$-$0.01 &$-$1.68 &$-$2.60 &$-$3.41 &$-$4.51 &$-$5.43 &$-$5.63 \\
$-$1.7 &  7.9 &  0.63 &  0.85 &  0.60 &  0.053\rlap{$^a$} &  1.09 &  0.10 &$-$1.50 &$-$2.39 &$-$3.17 &$-$4.24 &$-$5.13 &$-$5.33 \\
$-$1.7 & 10.0 &  0.64 &  0.86 &  0.59 &  0.054 &  1.17 &  0.24 &$-$1.28 &$-$2.13 &$-$2.88 &$-$3.92 &$-$4.78 &$-$4.97 \\
$-$1.7 & 11.2 &  0.65 &  0.86 &  0.59 &  0.056 &  1.19 &  0.29 &$-$1.19 &$-$2.02 &$-$2.76 &$-$3.77 &$-$4.62 &$-$4.81 \\
$-$1.7 & 12.6 &  0.65 &  0.86 &  0.58 &  0.058 &  1.22 &  0.36 &$-$1.09 &$-$1.90 &$-$2.62 &$-$3.62 &$-$4.45 &$-$4.63 \\
$-$1.7 & 14.1 &  0.65 &  0.86 &  0.58 &  0.055 &  1.25 &  0.43 &$-$0.98 &$-$1.77 &$-$2.47 &$-$3.45 &$-$4.27 &$-$4.44 \\
$-$1.7 & 15.8 &  0.64 &  0.85 &  0.58 &  0.058 &  1.27 &  0.50 &$-$0.86 &$-$1.64 &$-$2.32 &$-$3.28 &$-$4.07 &$-$4.24 \\
$-$1.7 & 17.8 &  0.63 &  0.85 &  0.58 &  0.060 &  1.27 &  0.55 &$-$0.78 &$-$1.53 &$-$2.20 &$-$3.13 &$-$3.90 &$-$4.06 \\
$-$1.3 &  4.0 &  0.62 &  0.88 &  0.68 &  0.074\rlap{$^a$} &  1.50 &  0.32 &$-$1.53 &$-$2.62 &$-$3.60 &$-$4.95 &$-$6.02 &$-$6.27 \\
$-$1.3 &  5.0 &  0.65 &  0.89 &  0.68 &  0.077\rlap{$^a$} &  1.59 &  0.42 &$-$1.39 &$-$2.44 &$-$3.41 &$-$4.76 &$-$5.82 &$-$6.07 \\
$-$1.3 &  6.3 &  0.67 &  0.90 &  0.67 &  0.079\rlap{$^a$} &  1.65 &  0.50 &$-$1.26 &$-$2.29 &$-$3.22 &$-$4.53 &$-$5.57 &$-$5.82 \\
$-$1.3 &  7.9 &  0.68 &  0.90 &  0.65 &  0.082\rlap{$^a$} &  1.74 &  0.61 &$-$1.09 &$-$2.08 &$-$2.97 &$-$4.23 &$-$5.25 &$-$5.49 \\
$-$1.3 & 10.0 &  0.70 &  0.91 &  0.65 &  0.083 &  1.81 &  0.71 &$-$0.93 &$-$1.88 &$-$2.73 &$-$3.94 &$-$4.92 &$-$5.15 \\
$-$1.3 & 11.2 &  0.71 &  0.91 &  0.64 &  0.082 &  1.82 &  0.75 &$-$0.87 &$-$1.80 &$-$2.62 &$-$3.79 &$-$4.76 &$-$4.98 \\
$-$1.3 & 12.6 &  0.71 &  0.92 &  0.64 &  0.086 &  1.84 &  0.79 &$-$0.81 &$-$1.71 &$-$2.52 &$-$3.65 &$-$4.60 &$-$4.81 \\
$-$1.3 & 14.1 &  0.72 &  0.92 &  0.63 &  0.088 &  1.84 &  0.82 &$-$0.75 &$-$1.64 &$-$2.43 &$-$3.54 &$-$4.46 &$-$4.67 \\
$-$1.3 & 15.8 &  0.71 &  0.92 &  0.63 &  0.093 &  1.91 &  0.92 &$-$0.61 &$-$1.48 &$-$2.24 &$-$3.31 &$-$4.21 &$-$4.40 \\
$-$1.3 & 17.8 &  0.70 &  0.91 &  0.63 &  0.094 &  1.83 &  0.96 &$-$0.53 &$-$1.38 &$-$2.13 &$-$3.18 &$-$4.06 &$-$4.24 \\
$-$0.7 &  4.0 &  0.73 &  0.99 &  0.77 &  0.124\rlap{$^a$} &  2.59 &  1.39 &$-$0.36 &$-$1.47 &$-$2.83 &$-$4.75 &$-$5.95 &$-$6.17 \\
$-$0.7 &  5.0 &  0.75 &  1.00 &  0.76 &  0.127\rlap{$^a$} &  2.72 &  1.49 &$-$0.26 &$-$1.35 &$-$2.68 &$-$4.57 &$-$5.76 &$-$5.98 \\
$-$0.7 &  6.3 &  0.77 &  1.01 &  0.76 &  0.133 &  2.85 &  1.59 &$-$0.15 &$-$1.23 &$-$2.53 &$-$4.38 &$-$5.57 &$-$5.79 \\
$-$0.7 &  7.9 &  0.79 &  1.03 &  0.76 &  0.144 &  2.95 &  1.68 &$-$0.05 &$-$1.13 &$-$2.40 &$-$4.21 &$-$5.39 &$-$5.61 \\
$-$0.7 & 10.0 &  0.82 &  1.04 &  0.76 &  0.155 &  3.00 &  1.74 &  0.03 &$-$1.04 &$-$2.28 &$-$4.02 &$-$5.19 &$-$5.42 \\
$-$0.7 & 11.2 &  0.83 &  1.05 &  0.76 &  0.161 &  3.00 &  1.76 &  0.06 &$-$0.99 &$-$2.22 &$-$3.91 &$-$5.07 &$-$5.31 \\
$-$0.7 & 12.6 &  0.84 &  1.06 &  0.76 &  0.167 &  2.99 &  1.78 &  0.09 &$-$0.95 &$-$2.15 &$-$3.78 &$-$4.93 &$-$5.18 \\
$-$0.7 & 14.1 &  0.85 &  1.07 &  0.76 &  0.172 &  2.96 &  1.80 &  0.13 &$-$0.91 &$-$2.09 &$-$3.66 &$-$4.80 &$-$5.06 \\
$-$0.7 & 15.8 &  0.86 &  1.07 &  0.76 &  0.177 &  2.91 &  1.83 &  0.18 &$-$0.85 &$-$2.00 &$-$3.50 &$-$4.64 &$-$4.90 \\
$-$0.7 & 17.8 &  0.86 &  1.07 &  0.75 &  0.183 &  2.80 &  1.83 &  0.21 &$-$0.80 &$-$1.92 &$-$3.33 &$-$4.44 &$-$4.72 \\
$-$0.4 &  4.0 &  0.80 &  1.05 &  0.84 &  0.158 &  3.27 &  1.97 &  0.24 &$-$0.83 &$-$2.26 &$-$4.38 &$-$5.58 &$-$6.06 \\
$-$0.4 &  5.0 &  0.82 &  1.07 &  0.84 &  0.167 &  3.43 &  2.07 &  0.35 &$-$0.72 &$-$2.14 &$-$4.22 &$-$5.41 &$-$5.88 \\
$-$0.4 &  6.3 &  0.84 &  1.09 &  0.84 &  0.177 &  3.56 &  2.15 &  0.43 &$-$0.63 &$-$2.05 &$-$4.05 &$-$5.24 &$-$5.69 \\
$-$0.4 &  7.9 &  0.86 &  1.10 &  0.83 &  0.186 &  3.68 &  2.26 &  0.55 &$-$0.51 &$-$1.93 &$-$3.90 &$-$5.09 &$-$5.54 \\
$-$0.4 & 10.0 &  0.88 &  1.12 &  0.83 &  0.197 &  3.74 &  2.32 &  0.63 &$-$0.42 &$-$1.84 &$-$3.77 &$-$4.96 &$-$5.38 \\
$-$0.4 & 11.2 &  0.89 &  1.13 &  0.84 &  0.201 &  3.74 &  2.35 &  0.68 &$-$0.37 &$-$1.79 &$-$3.73 &$-$4.91 &$-$5.34 \\
$-$0.4 & 12.6 &  0.90 &  1.14 &  0.84 &  0.205 &  3.73 &  2.38 &  0.72 &$-$0.32 &$-$1.74 &$-$3.69 &$-$4.87 &$-$5.30 \\
$-$0.4 & 14.1 &  0.91 &  1.16 &  0.84 &  0.212 &  3.72 &  2.40 &  0.76 &$-$0.28 &$-$1.70 &$-$3.64 &$-$4.82 &$-$5.25 \\
$-$0.4 & 15.8 &  0.92 &  1.17 &  0.84 &  0.219 &  3.71 &  2.44 &  0.80 &$-$0.23 &$-$1.65 &$-$3.55 &$-$4.72 &$-$5.13 \\
$-$0.4 & 17.8 &  0.93 &  1.18 &  0.84 &  0.223 &  3.46 &  2.38 &  0.80 &$-$0.22 &$-$1.64 &$-$3.48 &$-$4.65 &$-$5.05 \\
\ppm0.0 &  4.0 &  0.89 &  1.13 &  0.89 &  0.218 &  4.10 &  2.68 &  0.99 &$-$0.04 &$-$1.53 &$-$4.09 &$-$5.29 &$-$5.80 \\
\ppm0.0 &  5.0 &  0.90 &  1.14 &  0.89 &  0.224 &  4.20 &  2.79 &  1.11 &  0.08 &$-$1.41 &$-$3.93 &$-$5.14 &$-$5.63 \\
\ppm0.0 &  6.3 &  0.91 &  1.15 &  0.89 &  0.232 &  4.41 &  2.93 &  1.25 &  0.23 &$-$1.25 &$-$3.79 &$-$4.99 &$-$5.48 \\
\ppm0.0 &  7.9 &  0.93 &  1.18 &  0.89 &  0.244 &  4.57 &  3.02 &  1.34 &  0.33 &$-$1.16 &$-$3.71 &$-$4.92 &$-$5.41 \\
\ppm0.0 & 10.0 &  0.96 &  1.20 &  0.90 &  0.258 &  4.70 &  3.09 &  1.42 &  0.42 &$-$1.05 &$-$3.61 &$-$4.81 &$-$5.30 \\
\ppm0.0 & 11.2 &  0.97 &  1.22 &  0.90 &  0.265 &  4.75 &  3.11 &  1.46 &  0.46 &$-$1.00 &$-$3.59 &$-$4.79 &$-$5.29 \\
\ppm0.0 & 12.6 &  0.99 &  1.24 &  0.91 &  0.273 &  4.80 &  3.14 &  1.49 &  0.50 &$-$0.96 &$-$3.56 &$-$4.76 &$-$5.26 \\
\ppm0.0 & 14.1 &  1.01 &  1.25 &  0.92 &  0.280 &  4.81 &  3.15 &  1.52 &  0.54 &$-$0.92 &$-$3.55 &$-$4.75 &$-$5.24 \\
\ppm0.0 & 15.8 &  1.02 &  1.26 &  0.92 &  0.287 &  4.83 &  3.17 &  1.56 &  0.59 &$-$0.86 &$-$3.53 &$-$4.72 &$-$5.22 \\
\ppm0.0 & 17.8 &  1.03 &  1.28 &  0.92 &  0.294 &  4.83 &  3.20 &  1.60 &  0.64 &$-$0.81 &$-$3.50 &$-$4.70 &$-$5.20 \\
$+$0.2 &  4.0 &  0.94 &  1.17 &  0.92 &  0.252 &  4.44 &  2.99 &  1.32 &  0.33 &$-$1.11 &$-$3.95 &$-$5.16 &$-$5.64 \\
$+$0.2 &  5.0 &  0.94 &  1.17 &  0.91 &  0.256 &  4.53 &  3.13 &  1.47 &  0.47 &$-$0.97 &$-$3.82 &$-$5.03 &$-$5.52 \\
$+$0.2 &  6.3 &  0.97 &  1.20 &  0.92 &  0.269 &  4.78 &  3.25 &  1.57 &  0.57 &$-$0.88 &$-$3.74 &$-$4.94 &$-$5.43 \\
$+$0.2 &  7.9 &  0.99 &  1.22 &  0.93 &  0.282 &  4.95 &  3.33 &  1.66 &  0.68 &$-$0.75 &$-$3.73 &$-$4.95 &$-$5.46 \\
$+$0.2 & 10.0 &  1.02 &  1.25 &  0.92 &  0.296 &  5.10 &  3.42 &  1.76 &  0.78 &$-$0.63 &$-$3.52 &$-$4.72 &$-$5.21 \\
$+$0.2 & 11.2 &  1.03 &  1.26 &  0.93 &  0.305 &  5.15 &  3.42 &  1.77 &  0.81 &$-$0.58 &$-$3.50 &$-$4.71 &$-$5.20 \\
$+$0.2 & 12.6 &  1.05 &  1.28 &  0.94 &  0.314 &  5.19 &  3.43 &  1.80 &  0.85 &$-$0.53 &$-$3.49 &$-$4.70 &$-$5.20 \\
$+$0.2 & 14.1 &  1.06 &  1.29 &  0.94 &  0.323 &  5.23 &  3.45 &  1.83 &  0.89 &$-$0.47 &$-$3.48 &$-$4.69 &$-$5.19 \\
$+$0.2 & 15.8 &  1.08 &  1.31 &  0.95 &  0.331 &  5.25 &  3.48 &  1.87 &  0.94 &$-$0.43 &$-$3.47 &$-$4.68 &$-$5.18 \\
$+$0.2 & 17.8 &  1.09 &  1.33 &  0.95 &  0.338 &  5.29 &  3.52 &  1.92 &  0.99 &$-$0.38 &$-$3.46 &$-$4.67 &$-$5.17 \\
\hline
\end{tabular}\end{minipage}
\end{table*}

\newpage

\begin{table*}
\centering
\begin{minipage}{100mm}
  \caption{Comparison of $I$-band SBF Zero Points}
\label{tab:izeros}
\def\midofcol#1{\hfill\hskip -20pt plus 1000pt#1\hskip -12pt plus 1000pt\hfill}
\newdimen\digitwidth
\setbox0=\hbox{\rm0}
\digitwidth=\wd0
\catcode`?=\active
\def?{\kern\digitwidth}
  \begin{tabular}{ccl}
\hline
?$\overline M^0_I\,$\rlap{\footnote[1]{$I$-band SBF zero point evaluated at
$\vi=1.15$ using a fixed slope of 4.5. Formal uncertainties
on the empirical numbers are $\sim\pm0.1$\,mag (see SBF-II and Ferrarese \etal).}} & 
Method\rlap{\footnote[2]{`Spiral' for the empirical direct SBF-Cepheid tie via spiral bulges;
`group' for the empirical SBF-Cepheid tie via galaxy groups;
`theory' for stellar population models.}}
& \midofcol{Source} \\
\hline                                                
$-$1.74 & spiral & SBF-II  \\
$-$1.79 & spiral & Ferrarese \etal\ 
(2000)\rlap{\footnote[3]{Ferrarese \etal\ used the same SBF and Cepheid
distances as SBF-II, but a different weighting scheme.}} \\
$-$1.54 & spiral & SBF-II, adjusted for NGC\,4258\rlap{\footnote[4]{Resulting 
zero point if the Cepheid scale is revised according to the Herrnstein \etal\ (1999)
NGC\,4258 maser distance.}} \\
$-$1.62 & group  & SBF-II \\
$-$1.42 & group  & SBF-II, adjusted for NGC\,4258\rlap{$^d$} 
\vspace{3pt}\\
$-$1.47 & theory & This work, Eq.~\ref{eq:ibar} \\
$-$1.81 & theory & SBF-I, Worthey (1994) models \\
$-$1.25 & theory & Worthey Web page, Padua
option\rlap{\footnote[5]{The `Padua isochrones' used here are
those of Bertelli et\,al.\ (1994).}}\\
$-$1.79 & theory & Liu \etal\ (2000)\\
\hline
\end{tabular}
\vspace{-10pt}
\end{minipage}
\end{table*}

\end{document}